%% file: sn-article.tex
\documentclass[reprint, onecolumn,superscriptaddress,sn-mathphys]{revtex4-2} 
\usepackage[utf8]{inputenc}
\usepackage{graphicx}
\usepackage{amsmath}
\usepackage{amssymb}
\usepackage[version=4]{mhchem}
\usepackage{siunitx}
\usepackage{tabularx}
\usepackage{multirow}
\usepackage{adjustbox}
\input{preamble}

\begin{document}

\title[Data-driven framework for input/output lookup table reduction]{Data-driven framework for input/output lookup tables reduction: Application to hypersonic flows in chemical non-equilibrium}

\author{Cl\'ement Scherding}
\email{Corresponding author: clement.scherding@dalembert.upmc.fr}
\affiliation{
 Institut Jean le Rond d'Alembert, Sorbonne University, France}%
 
 \author{Georgios Rigas}
 \affiliation{Department of Aeronautics, Imperial College London, UK}
 
 \author{Denis Sipp}
\affiliation{DAAA, Onera, France}

\author{Peter J. Schmid}
\affiliation{Department of Mechanical Engineering, KAUST, SA}

\author{Taraneh Sayadi}
\affiliation{
 Institut Jean le Rond d'Alembert, Sorbonne University, France}%
 \affiliation{
 Institute for Combustion Technology, Aachen University, Germany}%

\begin{abstract}
\vspace{0.2cm}
Hypersonic flows are of great interest in a wide range of aerospace applications and are a critical component of many technological advances. Accurate simulations of these flows in thermodynamic (non)-equilibrium (accounting for high temperature effects) rely on detailed thermochemical gas models. While accurately capturing the underlying aerothermochemistry, these models dramatically increase the cost of such calculations. In this paper, we present a novel model-agnostic machine-learning technique to extract a reduced thermochemical model of a gas mixture from a library. A first simulation gathers all relevant thermodynamic states and the corresponding gas properties via a given model. The states are embedded in a low-dimensional space and clustered to identify regions with different levels of thermochemical (non)-equilibrium. Then, a surrogate surface from the reduced cluster-space to the output space is generated using radial-basis-function networks.
The method is validated and benchmarked on simulations of a hypersonic flat-plate boundary layer and shock-wave boundary layer interaction with finite-rate chemistry. The gas properties of the reactive air mixture are initially modeled using the open-source \acrshort{Mutation++} library. Substituting \acrshort{Mutation++} with the light-weight, machine-learned alternative improves the performance of the solver by up to 70\% while maintaining overall accuracy in both cases.
\end{abstract}


\maketitle

%---------------------------------------------------------------
\section{Introduction}\label{intro}

Chemical non-equilibrium effects have been shown to play an important role in the accurate simulation of flows at hypersonic conditions and in the computation of design characteristics, such as transition location or thermal loading \cite{Marxen2013,Marxen2014,Candler2019, direnzo2021}. Recent studies have identified these effects as causes of order-one changes in growth rates, response behavior or sensitivities, even though the corresponding variations in first-order flow statistics have been modest. These findings have in turn prompted a significant endeavor of augmenting existing flow solvers with non-equilibrium modules to account for finite-rate aerothermochemical features.
Simulations in this parameter regime introduce and
track a range of species in their inert or ionized
forms \cite{zhong2012direct,panesi2014nonequilibrium,paredes2019nonmodal}. Complementing the hydrodynamic state
vector by chemical components is a well-established
technique, for example in combustion or atmospheric simulations, but the required modeling of the inter-species interactions, such as dissociation, reaction and recombination \cite{Anderson2019}, for hypersonic applications poses great challenges. 
Much of this modeling is accomplished by lookup libraries, which act as repositories of tabulated chemical reactions encountered for a given flow state \cite{Scoggins2020}. When passing state-vector components to the library, amplitudes and time-scales for various forcing terms are returned, appearing as exogeneous inputs to the momentum, energy and species transport equations.

\begin{figure}
    \centering
    \includegraphics[width=0.80\textwidth]{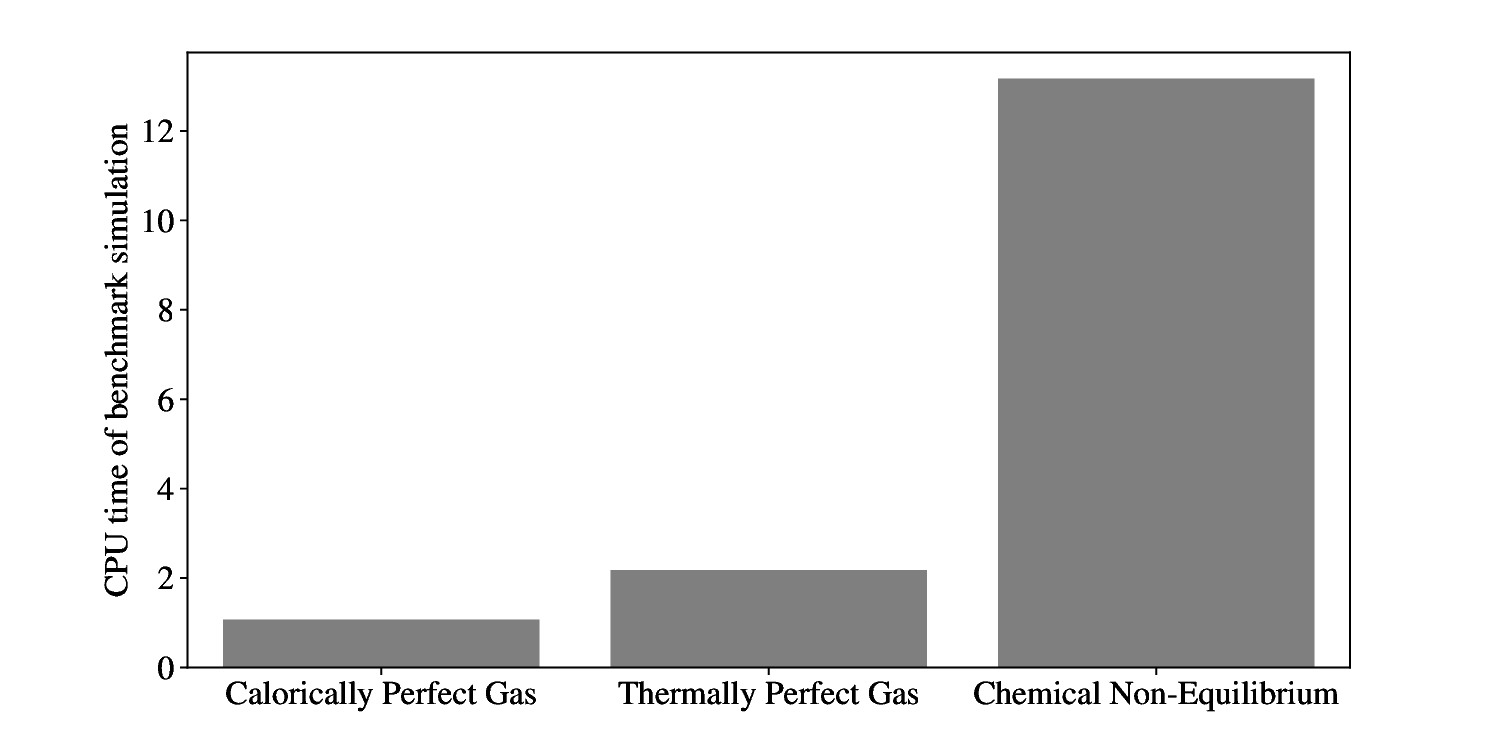}
    \caption{CPU time of a benchmark simulation, run with different aerothermochemical models. Including non-equilibrium effects in the simulation causes a significant increase in computational time.}
    \label{fig:CPU_cost}
\end{figure}

Much effort has gone into these libraries such as Pegase \cite{bottin1999thermodynamic}, Eglib \cite{ern2004eglib}, Plato \cite{munafo2020computational} and the leading library for reacting flows simulations CHEMKIN \cite{kee2000chemkin}. For aerothermochemical non-equilibrium effects in hypersonic flows, the \acrshort{Mutation++} library (MUlticomponent Thermodynamic And Transport properties for IONized gases in C++), developed and maintained at the von Karman Institute (VKI), has become the standard for high-fidelity simulations of high-speed and high-enthalpy flows \cite{Scoggins2020}. This library can be coupled to existing flow solvers and is capable of modeling a range of partially ionized gas effects, together with non-equilibrium features, energy exchange processes and gas-surface interactions. The flexibility and scope of the library comes at the expense of a computational bottleneck that slows down a typical large-scale simulation by a large factor, as shown in \cref{fig:CPU_cost}, where typical simulation times for calorically and thermally perfect gases are juxtaposed with results for non-equilibrium chemical reactions. A wide margin can be observed. For this reason, non-equilibrium computations range among the most inefficient and laborious calculations in fundamental hypersonic research. To increase performance, most CFD codes use hard-coded chemistry~\cite{di2020htr}. However, any change in the gas mixture or the thermochemical model comes at a human cost in terms of development, implementation and validation.\newline

More generally, many engineering applications need to evaluate an expensive function $\Tilde{f}(\mathbf{x})$ many times. Therefore, it is of great interest to alleviate the CPU burden of these applications by finding an efficient approximation of such functional forms. 
One of the oldest and most common approach to approximate $\Tilde{f}$ is to use structured tabulation. In a pre-processing step, values of $\Tilde{f}$ are tabulated for a hypercube in the input space. Then, during the simulation, values of $\Tilde{f}$ are linearly interpolated in the table. The Look-up Table method (LuT) was proven successful in many applications, such as tabulated chemistry for spray combustion \cite{franzelli2013tabulated}, design of energy devices using organic Rankine Cycles \cite{pini2015LUT} or simulations of hypersonic boundary layers in chemical equilibrium \cite{Marxen2011}. However, building and storing the table, together with the look-up procedure during the simulation, become computationally more intensive as the number of dimensions $D$ of the input space increases. This demonstrates the well-known curse of dimensionality, where the volume of sample points needed to construct an accurate table increases exponentially with the number of dimensions of the input space. Similarly, linear interpolation in high dimensions is a tedious task. This latter point even prevents the application of this LuT-methodology in the case considered in this study where the input space dimension is $D=6$. Pope developed the ISAT algorithm (\textit{in situ adaptive tabulation}) to overcome this deficiency in high dimensions with a storage/retrieval approach and demonstrate the concept on  applications in the combustion field \cite{pope1997computationally}. 

Recently, more general methods that can tackle higher dimensional problems have also been proposed and saw considerable success in a variety of applications, particularly in the active research field known as surrogate modeling. Underlying this effort is the universal approximation theorem \cite{cybenko1989approximation} which proves that deep neural networks, with at least one hidden layer and non-linear activation functions, formally proposed by LeCun \cite{lecun1985}, can approximate any non-linear function of any dimension. For example, Liu \textit{et al.} \cite{liu2000gradient} used neural-networks for surrogate-model-based optimization in aeronautics. However, the training cost of the network by back propagation becomes prohibitive as the number of neurons and layers increase -- a necessity which might arise in complex high-dimensional problems. Radial basis function (RBF) networks, a special case of three layers neural networks \cite{broomhead1988radial, buhmann2000radial}, can also be used for nonlinear function approximation in any dimension. Their training is easier and cheaper than classical neural networks as the optimal weights can be found by solving a linear system of equations. RBFs have been widely used for surrogate modeling in many fields such as aerodynamic shape optimization \cite{jin2001,peter2007} and meteorology \cite{chang2001}, to name but two.
Statistical surrogate modeling techniques have also found great success as they directly include an estimation of the error in the model. The method of kriging, originally developed for two-dimensional geostatistics problems \cite{krige1951statistical}, has been extended to approximate input/ouput problems of any dimension by Sacks \textit{et al.} \cite{sacks1989design}; see the review by Kleijnen on the use of kriging for surrogate modeling \cite{kleijnen2009kriging}. Finally, Polynomial Chaos Expansion (PCE) is another technique that can generate surrogate models well suited for uncertainty quantification  \cite{soize2004physical}. 

Despite some success, the often brute-force nature of these algorithms may not always yield a satisfactory surrogate model in terms of accuracy and computational cost. Bouhlel \textit{et al.} \citep{bouhlel2016} pointed out several performance issues when performing kriging in high dimensions ($D=100$). This number of dimensions is common in reactive flow simulations where hundreds of species are tracked, even with reduced chemical mechanisms \cite{attili2014formation,bansal2015direct,bhagatwala2014direct}. Moreover, one common assumption in surrogate modeling relates to the smoothness of the approximated relation. This is not always true, especially in hypersonic applications where shocks and temperature discontinuities are amongst the typical features of such flows. 
Nonetheless, clever pre-processing steps can greatly improve the model's performance in these cases. For example, Bouhlel et al. \cite{bouhlel2016} coupled kriging with partial least-squares (PLS) methods to reduce the high-dimensional ($D=100$) input space. In \cite{hawchar2017principal}, principal component analysis (PCA) has been used as a pre-processing step before applying polynomial chaos expansions on the PCA basis \cite{hawchar2017principal}. 
When dealing with discontinuous functions, Bettebghor \textit{et al.} \cite{bettebghor2011} proposed to cluster the input basis into different regions (to avoid a discontinuity within a cluster) and build a surrogate model on each of these regions. All models are then combined together and form a mixture of experts (MoE), as described in the literature \cite{hastie2009}. Yang \cite{yang2003regression}, however, pointed out that combining surrogate surfaces does not necessarily outperform a single model fitted over the entire input space. Hence, special care has to be taken in combining these steps.

The objective of this work is therefore to develop an effective pre-processing technique, allowing the construction of a low-dimensional surrogate model capable of replacing the computationally expensive library and the memory-intensive look-up tables when modeling inter-species interactions in simulations of chemically reactive flows.    

The paper is organized as follows. In \cref{algorithm}, the generic algorithm is presented in detail. It combines techniques from nonlinear model reduction, network clustering and surrogate modeling to efficiently extract a surrogate, light-weight model of the full library. In \cref{hypersonicflows}, the governing equations for hypersonic flows in chemical non-equilibrium as well as the thermo-chemical model for such flows are recalled. In \cref{results}, the algorithm is  first tested on the simulation of an adiabatic Mach-$10$ boundary layer in chemical non-equilibrium, initially studied by Marxen \textit{et al} \cite{Marxen2013}. The gas properties of the reactive air mixture are initially modeled using the open-source \acrshort{Mutation++} library \cite{Scoggins2020}. Replacing the library by the surrogate model then overcomes the computational bottleneck alluded to above. In addition, a shock-wave boundary layer interaction case is also considered in order to highlight the capability of the algorithm to deal with flows presenting discontinuities.

While focusing on the non-equilibrium gas-dynamic library \acrshort{Mutation++}, we stress that the employed techniques are agnostic about the particular library they are applied to, and can just as readily be employed to other libraries or lookup tables attached to simulations. Applications in combustion, phase-change simulations or particle-laden flows stand to benefit from this accelerating methodology at the interface between flow solvers and material-property libraries.

%-----------------------------------------------
\section{Description of the algorithm}\label{algorithm}

In this section, the general algorithm to extract a reduced library for the thermochemical properties of multi-component mixtures in chemical non-equilibrium is described. \\

Compressible flow simulations in chemical non-equilibrium require transport, thermodynamic and chemical reactions' properties, $\tilde{\mathbf{z}} \in \mathbb{R}^{D_Z}$, (\textit{e.g.} viscosity, conductivity, enthalpies, and chemical source terms) to close the governing equations. These properties are modeled as functions of the local thermodynamic state and mixture composition, concatenated into the local thermodynamic vector $\Tilde{\mathbf{q}}_{\rm{th}} \in \mathbb{R}^{D}$, usually computed using tabulation or external libraries, and can be considered as an input/output problem:
\begin{equation}
    \mathbf{\Tilde{z}} = \Tilde{f}(\Tilde{\mathbf{q}}_{\rm{th}}),
\end{equation}
where the function $\Tilde{f}$ represents the library of interest, and
$ D $ and $ D_Z $ represent the dimensions of the input and output spaces, respectively. This function then needs to be evaluated at each grid point and at each time-step. While accurate, the extensive calls to the library come at a substantial performance loss for the solver, together with a significant time penalty (see \cref{fig:CPU_cost}). \\

While these function calls cannot be entirely avoided, existing features of the flow inspire strategies to seek a less expensive method to evaluate the required properties. (i) Flows have history. In other words, several calls to the library may be redundant since some thermodynamic states are seen multiple times throughout the simulation. (ii) Any flow of interest contains only a subset of all possible thermodynamic states, given its nature and freestream conditions. Hence, only a small subset of the input space of function $\Tilde{f}$ needs to be accessed. (iii) While data-driven method requires a lot of data for training, some (rare) flows of interest, such as hypersonic boundary layers in chemical non-equilibrium, exhibit elegant, locally self-similar solutions \cite{lees1956} that can be used for training instead of an expensive direct numerical simulation (DNS). This final point will be explored in more detail in future work.  

\begin{figure}
    \centering
    \includegraphics[width=0.6\textwidth]{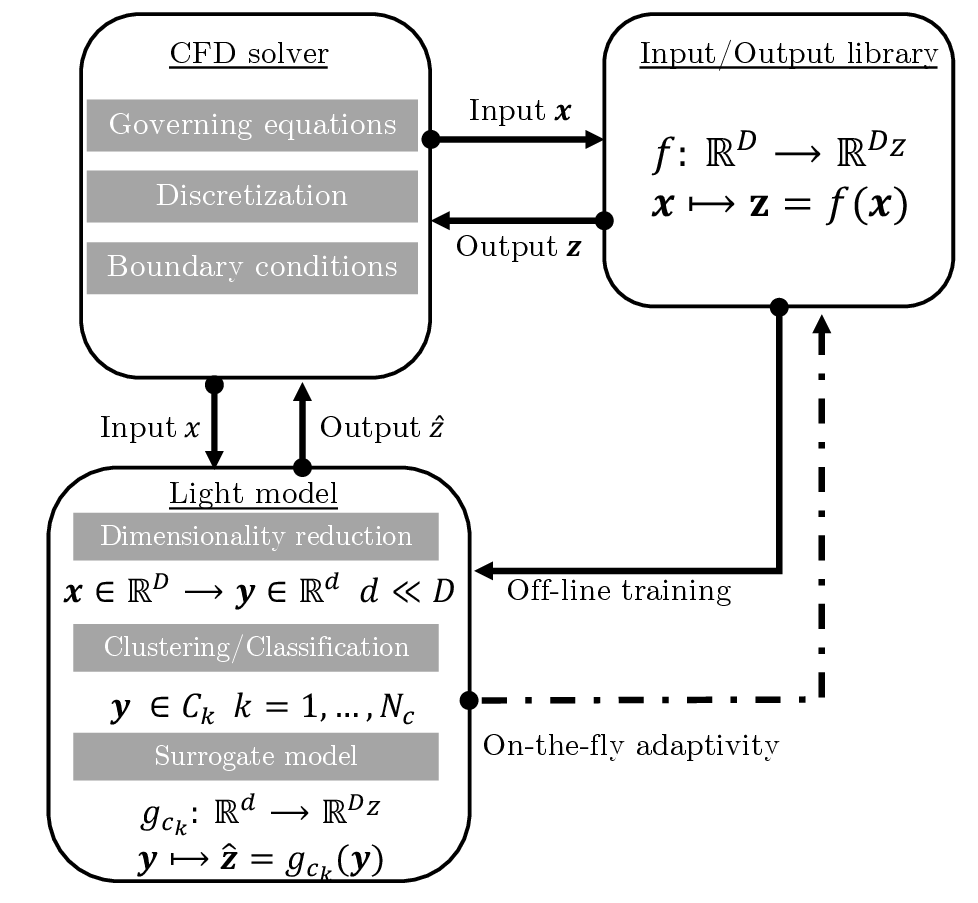}
    \caption{General schematic of the model training and coupling to replace any expensive Input/Output library.} 
    \label{fig:algo}
\end{figure}

The proposed algorithm leverages these features by creating a surrogate model of the function $\Tilde{f}$ only on a subset of input states relevant to the simulation, which is commonly represented as a low-dimensional manifold in $\mathbb{R}^{D}$. This allows us to first perform dimensionality reduction of the input space (see \cite{bouhlel2016,hawchar2017principal}). Next, following a similar approach as in \cite{bettebghor2011}, regions with different dynamics and/or discontinuities between them are clustered into a low-dimensional representation.
Finally, surrogate models are constructed on each cluster in this low-dimensional space. Hence, the training of the algorithm is performed in three steps: (i) dimensionality reduction, (ii) community clustering, and (iii) surrogate model construction. Once trained, the model replaces the look-up library already in place to predict the thermochemical properties of the mixture within the flow solver. We stress that the lighter version of the library will perform correctly only on the range of conditions seen during the simulation. A general schematic of the training process and the coupling with the flow solver is presented in \cref{fig:algo}. 

It should be noted that this strategy is also applicable to flows in thermal non-equilibrium, where the internal energy modes are out of equilibrium with the translational energy of the flow~\cite{munafo2015modeling}. Since the additional source terms modeling the energy exchange in the internal energy equation(s) are modeled as a function of the local thermodynamic state vector as well, they can be added to the outputs and a surrogate can be constructed accordingly. This endeavor however lies outside of the scope of the current study and will be pursued as a subject of a future work.
In the future, on-the-fly adaptivity will also be added to the algorithm to allow the model to learn new states as the simulation is advanced in time. This capability will help tackle more challenging and unsteady flow problems, while alleviating the need for a complete training set which may not always be available. Hence, the final intended use of the model will be as follows: (i) a preliminary laminar simulation will warm-start the training of the algorithm. (ii) as more flow features are added (such as instabilities), the model will learn adaptively the remaining ``new'' thermodynamics states pertaining to the new unsteady features. Following this procedure, the dynamics will be obtained at a lower CPU cost.

%-----------------------------------
\subsection{Training}
The algorithm is trained using the simulation of an adiabatic Mach-10 boundary layer in chemical non-equilibrium, thoroughly described in \cref{simu}. The gas considered is a five-species air model $\speciesset = { N_2, O_2, NO, N, O}$ with five reactions. The computational code solves the compressible reactive Navier-Stokes equations, where the kinetic parameters are computed by coupling with \acrshort{Mutation++} \cite{Scoggins2020}. The input thermodynamic state vector $\tilde{\textbf{q}}_{\rm{th}}$ is composed of density $\rho$, internal energy $\rho e$ and the mixture partial densities $\rho_s, s \in \speciesset$. The outputs of the library fall within three categories: thermodynamic properties (pressure $p$, temperature $T$, species specific enthalpies $h_s, s \in \speciesset$), transport properties (viscosity $\mu$, thermal conductivity $\kappa$, diffusion coefficients $D_s, s \in \speciesset$) and chemical kinetics source term $\dot{\omega}_s, s \in \speciesset$. More details concerning the governing equations and the thermo-chemical model can be found in \cref{hypersonicflows}. \\

\subsubsection{Data collection}
\label{DC}
\begin{figure}
    \centering
    \includegraphics[width=0.71\textwidth]{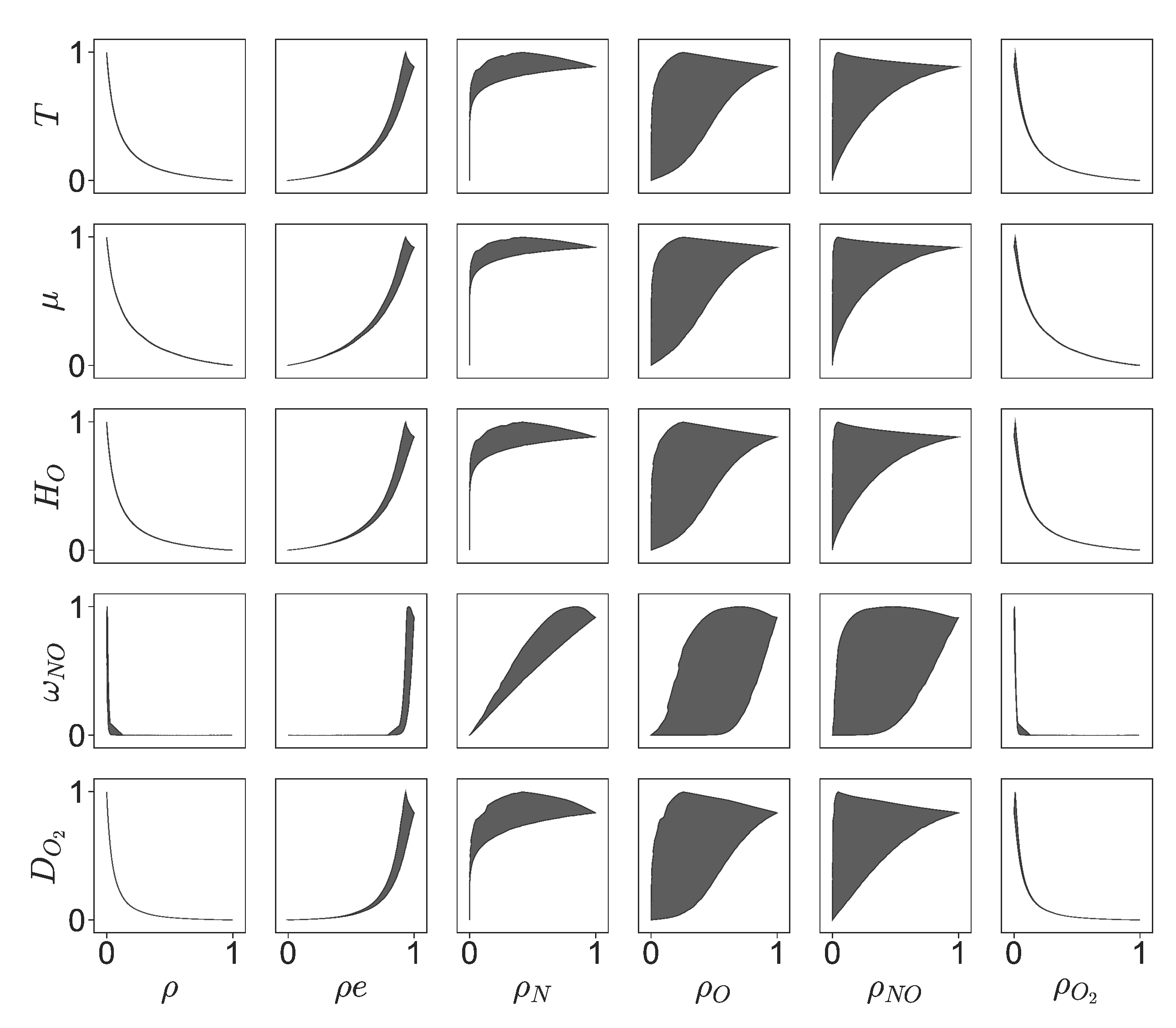}
    \caption{Numerical range of selected \acrshort{Mutation++} outputs $\mathbf{Z}$ (vertical) with respect to the inputs $\mathbf{X}$ (horizontal).}
    \label{fig:mpp_IO}
\end{figure}

To train the model, $N$ thermodynamic state vectors $\tilde{\textbf{q}}_{\rm{th}}$ are randomly sampled on the grid of a previously converged simulation in chemical non-equilibrium and concatenated into the input vector $\Tilde{\mathbf{X}} \in \mathbb{R}^{N \times D}$. The corresponding outputs from the library are collected and concatenated into the output vector $\Tilde{\mathbf{Z}} \in \mathbb{R}^{N \times D_Z}$. \cref{fig:mpp_IO} shows the numerical range of selected output variables along each input, normalized between $0$ and $1$ with a minimum-maximum scaling 
\begin{equation}
        \mathbf{X} = \frac{\Tilde{\mathbf{X}}-\Tilde{\mathbf{X}}_{min}}{\Tilde{\mathbf{X}}_{max}-\Tilde{\mathbf{X}}_{min}}, \quad \quad \mathbf{Z} = \frac{\Tilde{\mathbf{Z}}-\Tilde{\mathbf{Z}}_{min}}{\Tilde{\mathbf{Z}}_{max}-\Tilde{\mathbf{Z}}_{min}}.
\end{equation} \\

Taking dynamic viscosity $\mu$, for example, it shows a strong dependency on density $\rho$ and internal energy $\rho e$ but a low variation with respect to the radicals' partial densities $\rho_O$, $\rho_N$ and $\rho_{NO}$. The same observations can be made for all other outputs. Hence, the variation of the function $\Tilde{f}$ with respect to the inputs can be accurately represented on a low-dimensional subspace of the inputs. This motivates the first step of the algorithm, namely, dimensionality reduction. 

%-----------------------------------
\subsubsection{Dimensionality reduction}\label{lle}

The goal of this section is to find an effective algorithm for dimensionality reduction of the input space in order to construct a mapping between its reduced-order representation and the output of the library
\begin{equation}
    \hat{\mathbf{Z}} = g(\mathbf{Y}),
\end{equation}
where $g$ is the approximation of the scaled library $f$ in the low-dimensional subspace of the inputs, $\mathbf{Y}$ is the reduced-order representation of an input $\mathbf{X}$ and $\hat{\mathbf{Z}}$ the prediction of the model. 
The benefit of this first pre-processing step is to maintain high accuracy of the surrogate model, while decreasing the overall cost of construction and evaluation. In fact, constructing a response surface faces the well-known curse of dimensionality; as the number of input dimensions increases, the cost of constructing an accurate surface increases exponentially. This approach was proven successful in \cite{bouhlel2016} where PLS was used in tandem with kriging to reduce the dimension of the input space. \\

\begin{figure}
    \centering

    \subfloat[\label{subfig:PCA_T}]{%
      \includegraphics[width=0.45\columnwidth]{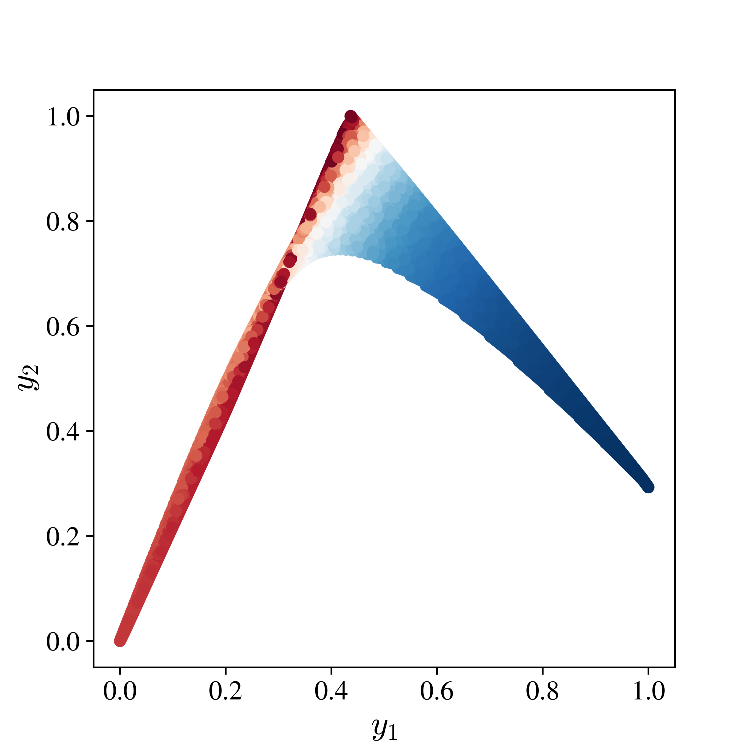}
    }
    \subfloat[\label{subfig:AE_T}]{%
      \includegraphics[width=0.45\columnwidth]{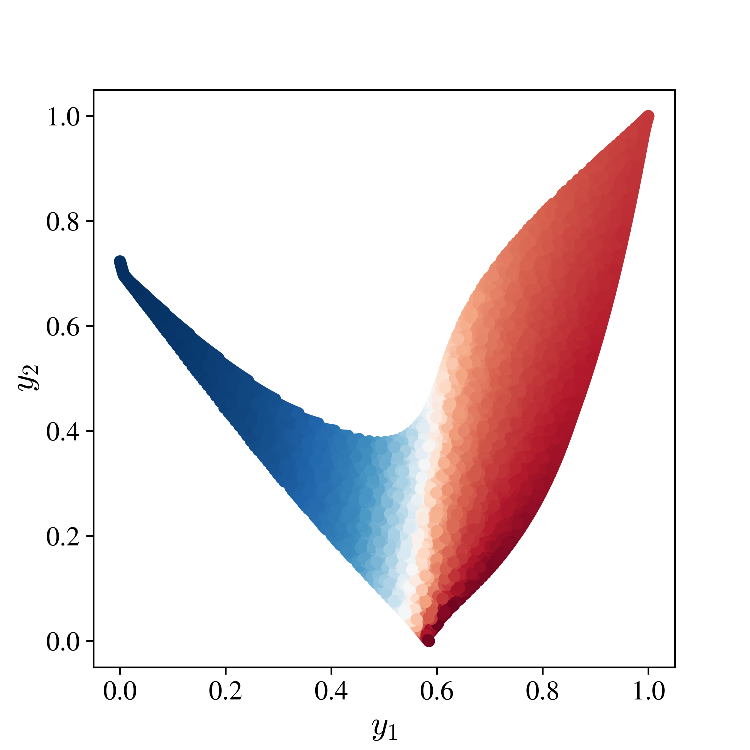}
    }
    
    \subfloat[\label{subfig:PLS_T}]{%
      \includegraphics[width=0.45\columnwidth]{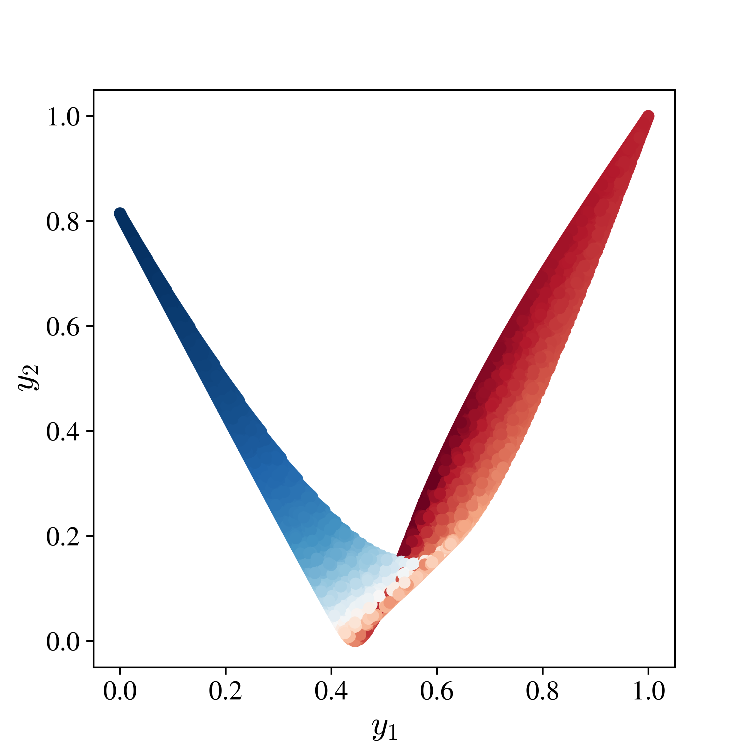}
    }
    \subfloat[\label{subfig:IOE_T}]{%
      \includegraphics[width=0.45\columnwidth]{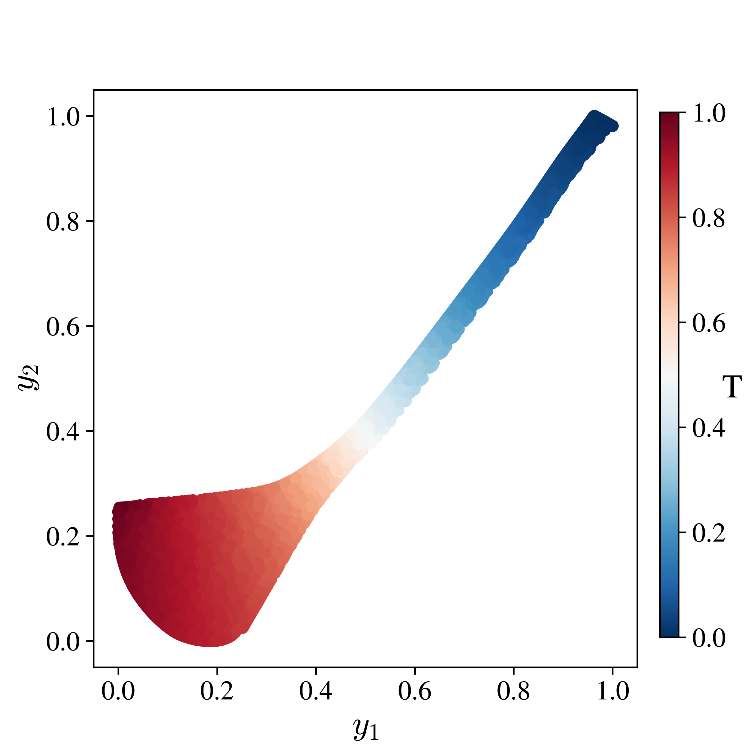}%
    }
    
    \caption{Low-dimensional representation $Y\in\mathbb{R}^{N \times d}$ ($d=2$) of $X\in\mathbb{R}^{N\times D}$, colored by temperature $T$.  Obtained with (a) PCA (b) AE (c) PLS (d) IO-E.}\label{fig:PCA_AE}
\end{figure}

\paragraph{Principal component analysis} The most common technique for dimensionality reduction of a dataset $\mathbf{X} \in \mathbb{R}^{N \times D} $ in high dimensions is principal component analysis PCA (see eg. \cite{shlens2014}). The principal components of $\mathbf{X}$ are found through the eigenvalue decomposition of the covariance matrix of the data. The dataset $\mathbf{X}$ is then projected onto $d < D$ leading eigenvectors (or principal components) of the covariance matrix, resulting in a low-dimensional representation $\mathbf{Y} \in \mathbb{R}^{N \times d}$ of the original dataset. However, depending on the shape of the manifold, the variations of the output variables with respect to the low-dimensional sub-space may not be properly preserved, which is the case presented in \cref{subfig:PCA_T} with points of high and low temperature projected onto similar locations. This example illustrates the limitations of PCA for dimensionality reduction of a dataset constrained to a nonlinear manifold. \\

%     \begin{subfloat}[b]{0.45\textwidth} 
%         \centering \includegraphics[width=\textwidth]{Figures/PCA_unrolling.eps}
%         \caption{}\label{fig:PCA_T}
%     \end{subfigure}
%     ~ 
%     \begin{subfigure}[b]{0.45\textwidth}
%         \centering \includegraphics[width=\textwidth]{Figures/LLE_unrolling.eps}
%         \caption{}\label{fig:AE_T}
%     \end{subfigure}
%     ~
%     \begin{subfigure}[b]{0.45\textwidth}
%         \centering \includegraphics[width=\textwidth]{Figures/PLS_unrolling.eps}
%         \caption{}\label{fig:PLS_T}
%     \end{subfigure}
%     ~
%     \begin{subfigure}[b]{0.45\textwidth}
%         \centering \includegraphics[width=\textwidth]{Figures/IO-E_unrolling.eps}
%         \caption{}\label{fig:IOE_T}
%     \end{subfigure}
%     \caption{Low-dimensional representation $Y\in\mathbb{R}^{N \times d}$ ($d=2$) of $X\in\mathbb{R}^{N\times D}$, colored by temperature $T$.  Obtained with (a) PCA (b) AE (c) PLS (d) IO-E.}\label{fig:PCA_AE}
% \end{figure}

\paragraph{Auto-encoders} Nonlinear dimensionality reduction via auto-encoders (AE) typically have a higher compression rate than linear techniques. An auto-encoder is a parametric model (i.e. a deep neural network with an activation function $\sigma$) that embeds the input dataset $\mathbf{X} \in \mathbb{R}^{N \times D} $ into a low-dimensional representation $\mathbf{Y} \in \mathbb{R}^{N \times d}$ through an encoder function $\mathbf{E}$. The low-dimensional representation is then decoded back to the input space with the decoder function $\mathbf{D}$, producing a reconstruction of the input $\widehat{\mathbf{X}} \in \mathbb{R}^{N \times D}$. 
\begin{equation}
    \begin{array}{l}
        \mathbf{Y} = \mathbf{E}(\mathbf{X}) \\
        \hat{\mathbf{X}} = \mathbf{D}(\mathbf{Y})
    \end{array}
\end{equation} 
The weights of the two networks $\mathbf{E}$ and $\mathbf{D}$ can be trained using back-propagation of the $L_2$ error $\lVert \mathbf{X} - \widehat{\mathbf{X}} \rVert_2$ through the network. If the activation function is selected as the identity (i.e $\sigma(\mathbf{x}) = \mathbf{x}$), the auto-encoder is linear and unbiased 
\begin{equation}
    \begin{array}{l}
        \mathbf{E} = W_E \\
        \mathbf{D} = W_D
    \end{array}
\end{equation} 
where $W_E$ and $W_D$ are the weights matrices of the encoder and decoder, respectively. These optimal weights can be found through PCA. In fact, the linear latent space of dimension $d$ of the encoder will span the same sub-space as the top $d$ PCA singular vectors. The equivalence between the two techniques was first shown by Baldi \& Hornik \cite{baldi1989neural}. Correspondingly, a two-layered nonlinear auto-encoder can be mathematically described as follows
\begin{equation}
    \begin{array}{l}
        \mathbf{Y} = W_{E,2} \sigma(W_{E,1}\mathbf{X}+\mathbf{b_{E,1}}) +\mathbf{b_{E,2}} \\
        \hat{\mathbf{X}} = W_{D,2} \sigma(W_{D,1}\mathbf{Y}+\mathbf{b_{D,1}}) +\mathbf{b_{D,2}}
    \end{array}
\end{equation}
where $\sigma$ is a nonlinear activation function, $W_{E,1} \in \mathbb{R}^{H \times D}, W_{E,2} \in \mathbb{R}^{d \times H}$ are the weights matrices of the first and second layer of the encoder with respective biases $\mathbf{b_{E,1}} \in \mathbb{R}^H$ and $\mathbf{b_{E,2}} \in \mathbb{R}^d$. $H$ denotes the dimension of the hidden layer. The matrices and bias vectors of the decoder have transposed dimensions. This corresponds to the minimal architecture (i.e. with one hidden nonlinear layer and an output linear layer) requested by the universal approximation theorem \cite{cybenko1989approximation}. However, more hidden layers can be considered. \cref{subfig:AE_T} shows the manifold unrolled in two dimensions with an auto-encoder, colored by the magnitude of the temperature, an output of the library. The AE outperforms PCA by preserving the local structure and preventing points at different thermodynamic states (i.e different temperatures) to be projected onto the same location. However, the highest temperature zone is concentrated in a thin layer adjacent to the lower temperature area. This will result in strong and unphysical gradients of the surrogate model within this region.\\

\paragraph{Partial least-squares} Since our interest lies in reducing the dimensionality of the input to construct a reduced-order surrogate model of the input/output relations, it is useful to entangle the input into a low-dimensional space that best reconstructs the outputs.
In analogy to PCA finding dependencies between the inputs, partial least-squares (PLS) finds a basis of the input space that optimally accounts for features in the output space. It has been used to construct surrogate models aimed at reducing the dimensions of the input space (see \cite{bouhlel2016}). Different variants of PLS now exist, using either a singular value decomposition (PLS-SVD) or iterative algorithms (such as PLS-W2A in \cite{wegelin2000survey}). While it has been shown that in cases where the dimension of the latent space is strictly greater than one, PLS-SVD differs from PLS-W2A and its variant PLS2, no major differences in the resulting latent space were observed. The results of PLS-SVD are presented here to highlight the similarity to PCA. Given the input $\mathbf{X}$ and output vectors $\mathbf{Z}$, the PLS-SVD algorithm \cite{wegelin2000survey} determines
\begin{equation}
    \mathbf{X}^T \mathbf{Z} = \mathbf{U \Sigma V}^T.
\end{equation}
Similar to PCA, the projection of the input is then obtained by projecting onto the $d<D$ top left singular vectors 
\begin{equation}
    \mathbf{Y} = \mathbf{X}\overline{\mathbf{U}}
\end{equation} 
where $\overline{\mathbf{U}} \in \mathbb{R}^{D \times d}$ is the truncated left-singular matrix. 
\cref{subfig:PLS_T} shows the training set projected onto the two-dimensional basis generated with PLS. As expected, adding the output information in the computation improves the output's representation in the reduced-order input basis compared to PCA. However, an artificial discontinuity is created near the high-gradient region that was not present in \cref{subfig:AE_T}. This highlights the lower compression rate of linear techniques compared to nonlinear ones. \\

\begin{figure}[htbp]
    \centering
    \includegraphics[width=\textwidth]{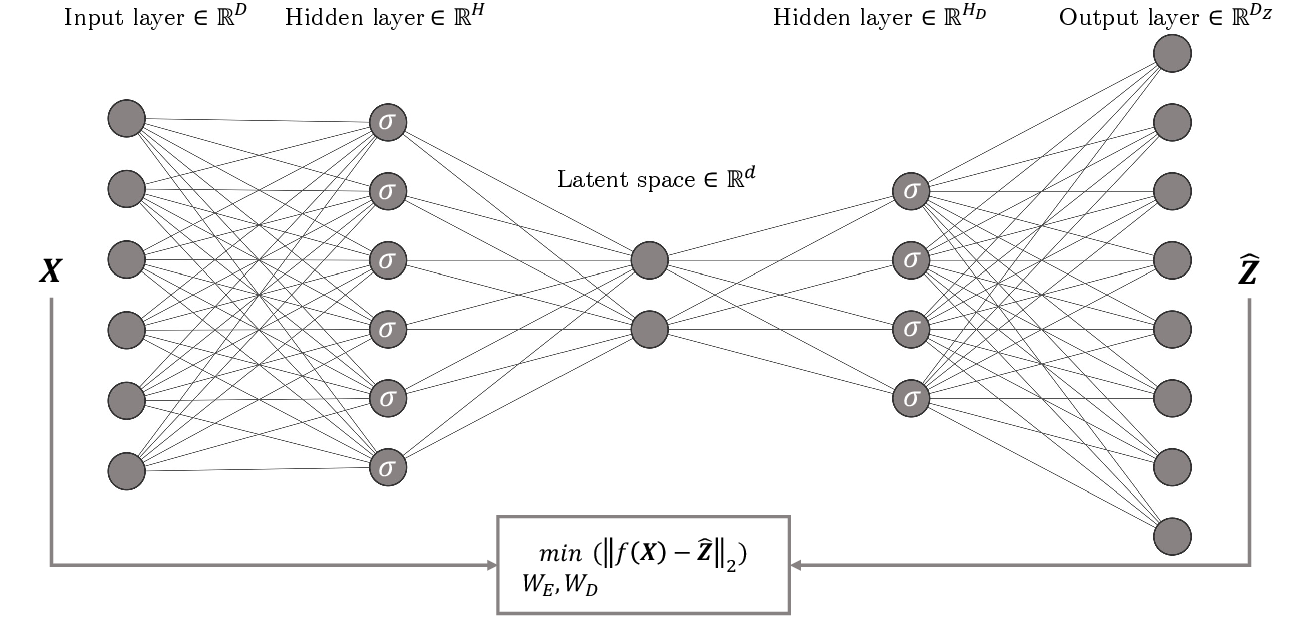}
    \caption{Schematic of the input/output-encoder architecture proposed. An input sample $\textbf{X}$ is passed through the encoder network to generate a low-dimensional representation $\textbf{Y}$. $\textbf{Y}$ is decoded back to predict the library output $\widehat{\textbf{Z}}$. At each training step, the $L_2$ loss $\lVert \mathbf{Z} - \widehat{\mathbf{Z}} \rVert_2$ between the library and the network prediction is calculcated and backpropagated until convergence of the encoder and decoder weights, $W_E$ and $W_D$, respectively, is reached.}
    \label{fig:encoder_architecture}
\end{figure}

\paragraph{Input/output-encoders} The strategy adopted here is therefore a nonlinear dimensionality reduction of the input data using input/ouput-encoders (IO-E), a modified version
of auto-encoders (AE) adapted to input/ouput relations. To that end, the decoder architecture is modified to best reconstruct the output of the library $\mathbf{Z} \in \mathbb{R}^{N\times D_Z}$ as 
\begin{equation}
    \widehat{\mathbf{Z}} = W_{D,2} \sigma(W_{D,1}\mathbf{Y}+\mathbf{b_{D,1}}) +\mathbf{b_{D,2}}
\end{equation}
where the weight matrices are now $W_{D,1} \in \mathbb{R}^{H_D \times d}, W_{D,2} \in \mathbb{R}^{D_Z \times H_D}$ with respective biases $\mathbf{b_{D,1}} \in \mathbb{R}^{H_D}$ and $\mathbf{b_{D,2}} \in \mathbb{R}^{D_Z}$. $H_D$ now represents the hidden dimension of the decoder. The resulting network is then trained via back-propagation of the reconstruction error based on the output $\lVert \mathbf{Z} - \widehat{\mathbf{Z}} \rVert_2$. The input/output-encoder (IO-E) architecture is illustrated in \cref{fig:encoder_architecture}. \\

\cref{subfig:IOE_T} shows the equivalent manifold using the IO-E architecture, with the same number of hidden dimensions. The IO-E architecture outperforms all other techniques as the high-temperature region is now projected onto its own properly defined zone with smooth variations. This feature will have a marked impact on the performance of the library when linked to the solver. It is worth noting that the decoder part of the IO-E could be used directly to predict the output of the library. However, we will see in \cref{mode_acc} that the accuracy of the full IO-E would not optimal in this case. This motivates the use of clustering and radial basis functions, as described next.

%-----------------------------------
\subsubsection{Community clustering}\label{clustering}
In the second step of the algorithm, we seek to discover clusters within our data. In our present context, a cluster represents a subset of data that shares similar thermodynamic features. These feature classification will then be useful in constructing a dedicated surrogate surface of the low-dimensional input manifold. To this end, Newman's spectral algorithm for community detection in a network \cite{newman2006} is used. A clear advantage of Newman's algorithm is that the number of clusters is not defined {\it{a priori}}, in contrast to more common clustering techniques (e.g. k-means). The number of clusters is instead the result of a maximization procedure performed on the modularity $Q$ of the network. In other words, the number of thermodynamic clusters in the flow are determined only from the data and is not based on {\it{a priori}} knowledge of the user. This knowledge might be even impossible to come by in complex, unsteady hypersonic flows subjected to shocks.

Following this approach, the Euclidean distance matrix $\Delta$ of the dataset in low-dimensional space is computed 
\begin{equation}
    \Delta_{ij} = \lVert \mathbf{Y}_i - \mathbf{Y}_j \rVert^2 \quad \text{for} \quad (\mathbf{Y}_i,\mathbf{Y}_j) \in \mathbf{Y}^2.
\end{equation}
The dataset is subsequently recast into an undirected network with a binary adjacency matrix $A$, constructed as
\begin{equation}
    A_{ij} = \left\{
    \begin{array}{lcl}
        1 &\phantom{1234}& \mbox{if} ~ \Delta_{ij}<\epsilon,  \\
        0 && \mbox{otherwise.}
    \end{array}
\right.
\end{equation}
Two points are connected with an edge if their Euclidean distance is below a certain threshold $\epsilon$. This threshold is usually chosen as a fraction of the mean of the distance matrix $\Delta$. The influence of the threshold $\epsilon$ on the number of clusters will be investigated in \cref{cluster-training}. 

Finally, the dataset is progressively split into two communities until the modularity, $Q$, is maximized. The modularity is defined as the proportion of edges contained within a community over the same proportion for a random reference network. Let $k_i$ be the number of edges pointing towards the data point numbered $i$, and $m$ the total number of connections within the network. Then, the probability of having an edge between $i$ and $j$ in the random reference network is $k_ik_j/m$. Hence, the modularity $Q$ is defined as,
\begin{equation}
    Q = \frac{1}{m}\sum_{ij}\left( A_{ij}-\frac{k_ik_j}{m}\right)\delta_{c_i,c_j},
\end{equation}
where $\delta_{c_i,c_j}$ is the Kronecker delta for the communities of $i$ and $j$. For instance, $\delta_{c_i,c_j}=1$ only when $i$ and $j$ belong to the same community. By defining a vector $\textbf{s}$ where $s_i=1$ if vertex $i$ belongs to the first group, and $s_i=-1$ otherwise, we can reformulate
\begin{equation}
    Q = \frac{1}{2m}\sum_{ij}\left( A_{ij}-\frac{k_ik_j}{m}\right)(s_is_j+1)=\frac{1}{2m}\mathbf{s^T} B \mathbf{s},
\end{equation}
where the modularity matrix $B$ is given as,
\begin{equation}
    B_{ij} = A_{ij}-\frac{k_ik_j}{m}.
\end{equation}
Since the graph is undirected, the modularity matrix $B$ is symmetric and the modularity $Q$ represents a Rayleigh quotient for matrix $B$. In order to maximize $Q$, we need to choose a vector $s$ that is parallel to the principal eigenvector (corresponding to the largest eigenvalue) of $B$, $\textbf{v}$, which can be achieved by setting $s_i=1$ if $v_i>0$ and $s_i=-1$ if $v_i<0$. In order to partition the graph into more than two communities, this algorithm is repeated until the modularity of each subgraph can no longer be increased. A thorough descripion of the full algorithm can be found in \cite{newman2006}.

\begin{figure}[htbp]
    \centering

    \subfloat[\label{subfig:adjacency_scrambled}]{%
      \includegraphics[width=0.45\columnwidth]{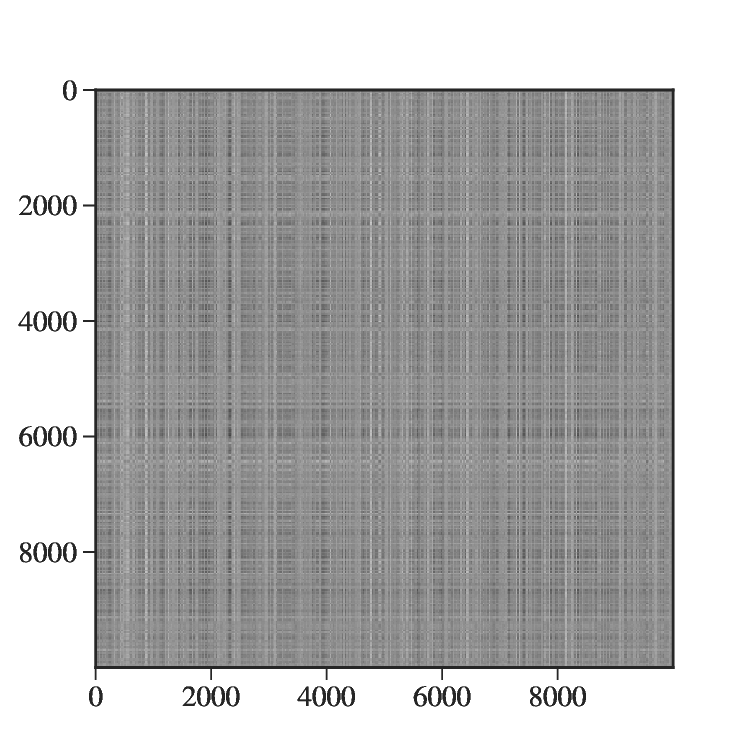}
    }
    \subfloat[\label{subfig:adjacency_cluster}]{%
      \includegraphics[width=0.45\columnwidth]{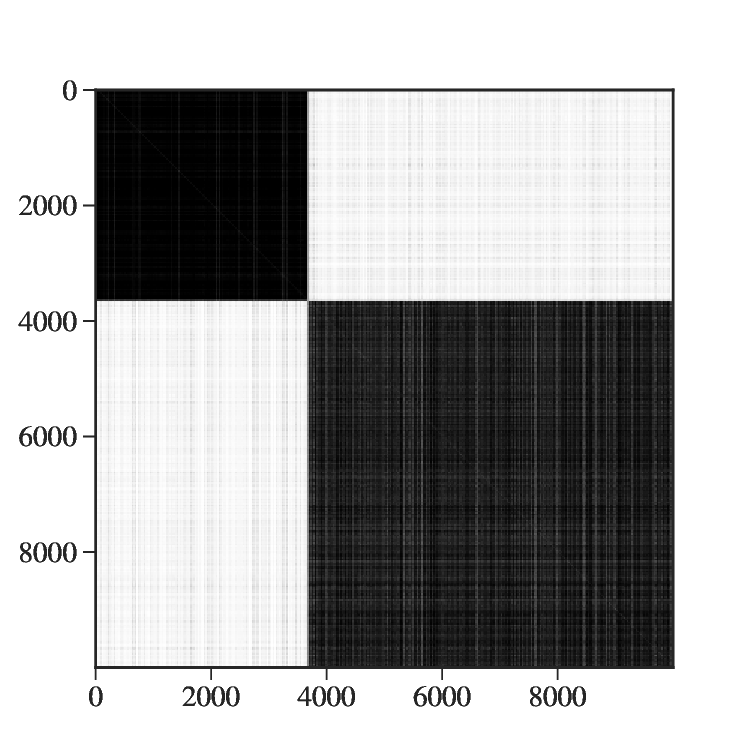}
    }
    \caption{Demonstration of the Newman algorithm. (a) Original distance matrix $\Delta$ of dataset $\mathbf{Y}$. (b) Restored communities from running Newman's algorithm on $A$, showing two distinct clusters}\label{fig:adjacency}
\end{figure}

% \begin{figure}[htbp]
%     \centering
%     \begin{subfigure}[b]{0.45\textwidth} 
%         \centering \includegraphics[width=\textwidth]{Figures/adjacency_scrambled.eps}
%         \caption{}\label{fig:adjacency_scrambled}
%     \end{subfigure}
%     ~ 
%     \begin{subfigure}[b]{0.45\textwidth}
%         \centering \includegraphics[width=\textwidth]{Figures/adjacency_clustered.eps}
%         \caption{}\label{fig:adjacency_cluster}
%     \end{subfigure}
%     \caption{Demonstration of the Newman algorithm. (a) Original distance matrix $\Delta$ of dataset $\mathbf{Y}$. (b) Restored communities from running Newman's algorithm on $A$, showing two distinct clusters}\label{fig:adjacency}
% \end{figure}

\cref{subfig:adjacency_scrambled} shows the application of the clustering algorithm to the boundary layer data. The distance matrix $\Delta$ is constructed on $\mathbf{Y}$. After running the algorithm on the subsequent adjacency matrix $A$, two distinct clusters are identified, highlighted by the low distance between the points within a cluster in \cref{subfig:adjacency_cluster}.

%-----------------------------------
\subsubsection{Surrogate model construction}\label{MR}
Finally, a surrogate surface is computed on the scattered low-dimensional points of each cluster. Many algorithms can be used to that end, such as kriging \cite{kleijnen2009kriging, bouhlel2016}, artificial neural networks \cite{sun2019review}, or radial basis functions \cite{broomhead1988radial,powell1992rbf, buhmann2000radial}. 
In the application of interest to this work, radial basis functions (RBF) provided the best trade-off between performance and accuracy, as well as an easy training step. In fact, the optimal weights of a RBF can be obtained through the solution of a linear system. However, it should be noted that our choice is not definitive and can be easily changed. 

\begin{figure}
    \centering
    \includegraphics[width=0.45\columnwidth]{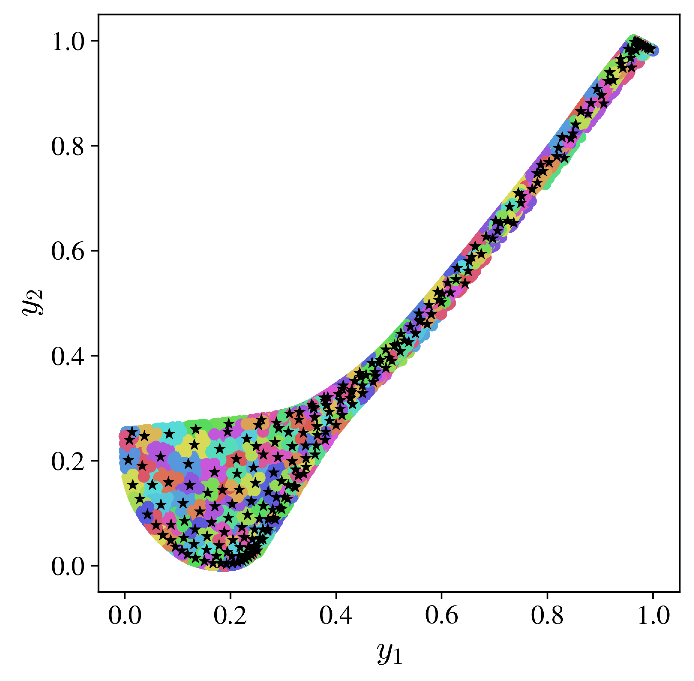}
    \caption{Tesselation of the low-dimensional space after applying the \textit{k-means} algorithm with $N_R = 250$ on the concatenation of the input $\textbf{X} \in \mathbb{R}^{N\times D}$ and output  $\textbf{Z} \in \mathbb{R}^{N\times D_Z}$ vectors. Black stars represent the cluster centroids' low-dimensional representation $\textbf{y}^c_{1},..., \textbf{y}^c_{N_R}$.}
    \label{fig:k-means}
\end{figure}

Given a set of $N_R$ input points $\textbf{x}_1, ..., \textbf{x}_{N_{R}} \in \mathbb{R}^d$ and the function value at these points $f(\textbf{x}_1), ..., f(\textbf{x}_{N_{R}})$ the radial basis function (RBF) interpolant $g$ is given by
\begin{equation}
\label{eq:rbf}
   g(\phi,x) = \sum_{i=1}^{N_R} \lambda_i \phi(\lVert \textbf{x} - \textbf{x}_i \rVert)
\end{equation}
where $\phi$ is the kernel function whose value depends on the distance $r = \lVert \textbf{x}-\textbf{x}_i \rVert$ between the evaluation point $\textbf{x}$ and the center $\textbf{x}_i$ of the RBF. In this study, the thin-plate spline kernel \cite{wood2003thin} was used, i.e., 
\begin{equation}
    \phi(r)=r^2 log(r).
\end{equation}
With this particular kernel, the kernel matrix is only conditionally positive definite. To ensure a unique solution for the interpolation weights, the system is augmented by a polynomial $ p \in \Pi^d_m $ (space of polynomials of $d$ variables and degree up to $m$) to the right hand side of \cref{eq:rbf} \cite{buhmann2000radial}, resulting in 
\begin{equation}
\label{eq:rbf_poly}
   g(x) = \sum_{i=1}^{N_R}  \lambda_i \phi(\lVert x - x_i \rVert) + p(x).
\end{equation}
The extra degrees of freedom are accounted for by enforcing orthogonality of the coefficients with respect to the polynomial space as 
\begin{equation}
\label{eq:poly_orthogonality}
  \sum_{i=1}^{N_R} \lambda_i r(\textbf{x}) = 0, r \in \Pi_m^d.
\end{equation}
Finally, the polynomial coefficients $\textbf{c} = [c_{1},..., c_{d+1}]^T$ and the RBF coefficients $\boldsymbol\Lambda = [\lambda_{1},..., \lambda_{N_{R}}]^T$ are found through the solution of the following linear system
\begin{equation}
    \begin{bmatrix} \mathbf{\Phi} & \mathbf{P} \\ \mathbf{P}^T & 0   \end{bmatrix}
 \begin{bmatrix} \boldsymbol\Lambda \\ \mathbf{c} \end{bmatrix} = 
    \begin{bmatrix} \mathbf{f} \\ 0^{d+1} \end{bmatrix}
\end{equation} \\ 
where $\textbf{P}_i = [1,x_{i},..., x_{i}^d]$ for $i \in [1, N_R]$ is the polynomial matrix, and $\textbf{f} = [f(\textbf{x}_1), ..., f(\textbf{x}_{N_{R}})]$ denotes the vector containing the function values at the RBF centers. \\ 
Due to the large size of the training set, using one cluster center per training point (i.e. $N_R = N = O(10^5)$) will likely result in overfitting and prohibitive computational cost \cite{schwenker2001three}. Following the recommendations in \cite{schwenker2001three}, the interpolant is constructed in two steps. First, the \textit{k-means} algorithm with $N_R \ll N$ clusters is applied on the concatenation of the input and output vector $(\mathbf{X},\mathbf{Z}) \in \mathbb{R}^{N \times (D+D_Z)}$. The addition of the output vector results in a low within-cluster variance of the outputs and will ultimately improve the surrogate model. The $N_R$ centroids obtained with \textit{k-means}, $\textbf{x}^c$,  are sent to the library to compute the function value vector $\textbf{f}$.  Simultaneously, they are encoded in the low-dimensional space to obtain the $N_R$ cluster centers that will be used to train the RBF $\textbf{y}^c_1, ..., \textbf{y}^c_{N_{R}} \in \mathbb{R}^d$. \cref{fig:k-means} shows the resulting tesselation in the embedded space after applying the \textit{k-means} algorithm with $N_R = 250$. The influence of the number of RBF centers on the quality of the surrogate model will be assessed in \cref{rbf-training}. \\

Following this approach, a single interpolant, $g_{c_k}$, is constructed for each cluster, $c_k$; in other words, $g_{c_k}$ is the approximation of the scaled library function $f$ on the low-dimensional subspace corresponding to cluster $c_k$. The advantage of having one interpolant per cluster is twofold. First, it allows the surrogate model to best fit a region with a given dynamics of the high-dimensional function, especially in the presence of discontinuities, similar to the approach of Bettebghor et al. \cite{bettebghor2011}. Secondly, as the surrogate model spans a smaller range of input parameters, less centers are required to capture the given dynamics accurately, resulting in a lighter model with faster evaluation time. 
To enforce continuity of the surrogate surface near the cluster boundaries, the nearest centroids that do not belong to the considered cluster are added to its training set. 

%-----------------------------------
\subsection{Coupling to the solver}
Once the model is trained, we can replace the calls of the solver to the look-up library with the new lighter model.  New points have to go through three steps:
(i) Out-of-sample dimensionality reduction, (ii) classification, and (iii) interpolation. These three steps will be described in the following. 
Let $\mathbf{X}^{t} \in \mathbb{R}^{N_t \times D}$ denote the stack of all new points. 
%-----------------------------------
\subsubsection{Out-of-sample encoding}
The low-dimensional representation of $\mathbf{X}^{t}$ (after proper scaling of $\tilde{\mathbf{X}}^{t}$) is straightforward. Indeed, the point is simply fed to the encoder portion of the input/output-encoder 
\begin{equation}
    \mathbf{Y}^t = W_{E,2} \sigma(W_{E,1}\mathbf{X}^{t}+\mathbf{b_{E,1}}) +\mathbf{b_{E,2}}.   
\end{equation}
This results in a fast and inexpensive encoding of the new out-of-sample points. In fact, the time complexity of the encoding step is $O(H \times C_{ac} \times L \times N_t )$, where $H$ is the maximum number of neurons in a layer, $C_{ac}$ is the complexity of the activation function and $L$ is the number of layers in the encoder step of the input/output-encoder. 

%-----------------------------------
\subsubsection{Classification}
\begin{figure}[htbp]
    \centering
    \includegraphics[width=0.35\textwidth]{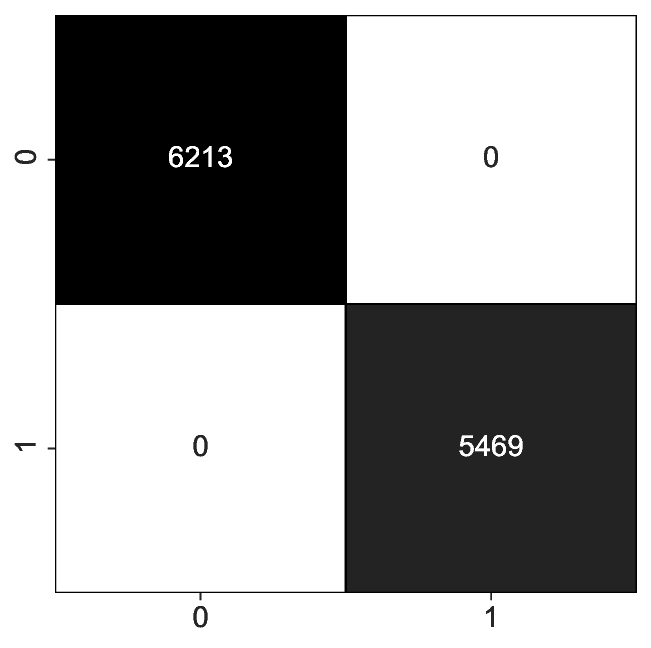}
    \caption{Confusion matrix $C$ of the random forest classifier ($n_{tree}=20$), tested on unseen data points during training. $C_{ij}$ is the number of points belonging to cluster $c_j$ predicted to be in cluster $c_i$.}
    \label{fig:confusion matrix}
\end{figure}

The next step is to determine to which community the new state $\mathbf{Y}^t$ belongs. To this end, after applying Newman's algorithm, a random forest classifier is trained on the resulting clusters. A random forest classifier, formally proposed by Breiman \cite{breiman2001random}, is a collection of $n_{tree}$ tree-based classifiers, ${h(\textbf{x},\Phi_k), k = 1,..., n_{tree}}$, where $\Phi_k$ are identically distributed random vectors. Each tree votes for the most likely class of input vector $\textbf{x},$ and the majority wins. 
\cref{fig:confusion matrix} shows the confusion matrix $C$ of the classifier trained on the two clusters obtained above with $n_{tree}=20$. The clusters of the embedded new points $\mathbf{Y}^{t}$, not seen during the training-phase of the classifier, are predicted and compared to the true clusters (given by Newman's algorithm). The off-diagonal values count the number of points that are assigned to the wrong cluster. All off-diagonal values are zero, demonstrating the ability of the classifier to correctly predict the community of an out-of-sample point.

The time complexity of the induction time of the classifier is $O(n_{tree} \times N_t log(N_t))$ (see the book by Witten, Frank and Hall \cite{frank2004data} for a demonstration), where $n_{tree}$ is the number of decision trees in the random forest. The logarithmic term accounts for the worst case scenario for the maximum depth of each tree. However, the maximum depth is usually set to a smaller value, resulting in a complexity of $O(n_{tree} \times depth \times N_t)$, which results in relatively fast classification times. 

%-----------------------------------
\subsubsection{Interpolation}
Finally, once $\mathbf{Y}^t$ has been found to belong to cluster $c_k$, the corresponding RBF $g_{c_k}$ is called to evaluate the thermochemical properties $\widehat{\mathbf{Z}}^{t}$ of the mixture at that state $\mathbf{X}^t$,
\begin{equation}
   \widehat{\mathbf{Z}}^{t} = g_{c_k}(\mathbf{Y}^t) = \sum_{i=1}^{N_R} \lambda_i \phi(\mathbf{Y}^t - \mathbf{Y}^c_i).
\end{equation}
The hat denotes the predicted value by the reduced library, as opposed to the true value that would have been given by the target library, $\mathbf{Z}^{t}$. Finally, these properties are re-scaled according to  
\begin{equation}
   \Tilde{\widehat{\mathbf{Z}}}^{t} =  \widehat{\mathbf{Z}}^{t} * (\Tilde{\mathbf{\mathbf{Z}}}_{max}-\Tilde{\mathbf{Z}}_{min}) +  \Tilde{\mathbf{Z}}_{min}
\end{equation}
and passed back to the flow solver.

The time complexity of each surrogate model is given by $O( C_{RBF} \times N_t \times N_R \times d)$, where $C_{RBF}$ is the complexity of the kernel function. The time limiting part of the RBF interpolation is the calculation of the distance matrix that scales with $O(N_t \times N_R \times d)$. Hence, using a relatively small number $N_R$ of RBF centers, compared to $N_t,$ will greatly improve the performance of the surrogate model. It should be noted that the dimensionality reduction directly contributes to the added performance of the surrogate model as $d<D$.

\subsubsection{Global performance}

The time complexity of the whole algorithm can be recovered by adding the time complexity of each of the three steps. Hence, the total time complexity of the algorithm can be written as $O(C_{ML}N)$, where $C_{ML} = O(HC_{ac}L + n_{tree}depth + d N_R C_{RBF})$. 

%-----------------------------------
\section{Application to hypersonic flows in chemical non-equilibrium}\label{hypersonicflows}
In this section, the general equations governing hypersonic flows in chemical non-equilibrium are first recalled, followed by a brief description of the numerical framework used to solve these equations. 
Then, the two benchmark cases, a Mach-10 adiabatic laminar boundary layer in chemical non-equilibrium, initially studied by Marxen et \textit{al} \cite{Marxen2013,Marxen2014}, and a Mach-5.92 shock wave boundary layer interaction are presented.

\subsection{Governing conservation equations}\label{sub:goveq}

The nondimensional Navier-Stokes equations for a mixture of multiple species \speciesset~ are presented in \crefrange{eq:globalmass}{eq:totenergy}.
\begin{alignat}{4}
	&\pd{\density}{\timevar} 
	&&+ \divgp{\density \velocity}
	&&= 0
% 	&&
    \label{eq:globalmass} \\
	\left\{\vphantom{\frac{1}{1}}\right.
	&\pd{\densityspecies}{\timevar} 
	&&+ \divgp{\densityspecies \velocity + \densityspecies \diffvelocity}
	&&= \massprod
	\left.\vphantom{\frac{1}{1}}\right\} \;,\; 
% 	&&
	\forall \; \speciess \in \speciesset
	\label{eq:speciesmass} \\
	&\pd{\density \velocity}{\timevar}
	&&+ \divgp{\density \velocity \otimes \velocity}
	&&= - \grad{\pressure} + \divg{\viscstresstensor} 
% 	&&
	\label{eq:momentum} \\
	&\pd{\density \totale}{\timevar}
	&&+ \divgp{\density \totalh \velocity}
	&&= \divgp{\viscstresstensor \cdot \velocity} - \divg{\heatflux} 
% 	&&
	\label{eq:totenergy}
\end{alignat}
\Cref{eq:globalmass} is the continuity equation, describing mass conservation in the system. \Cref{eq:speciesmass} corresponds to the set of mass conservation equations for each species, with the net production rate terms, \massprod, appearing on their right-hand side. In order to ensure global mass conservation, in the case of a finite-rate reacting mixture with a varying composition, \cref{eq:globalmass} needs to be solved together with \cref{eq:speciesmass} for all but one species. The omitted species is selected based on numerical considerations, avoiding species with the smallest concentrations.

The nondimensional quantities are the time, \timevar, the density, \density, the velocity, \velocity, the pressure, \pressure, the stress tensor, \viscstresstensor, the total energy, \totale, the total enthalpy, \totalh, the heat flux, \heatflux, as well as, the partial density, \densityspecies, the net mass production rate, \massprod, and the diffusion velocity, \diffvelocity, for the species \speciess. More details regarding the derivation of the equations and the validity of our invoked assumptions are provided in \cite{Anderson2019,Gnoffo1989,Josyula2015_equations}. 
There are $\numspecies - 1 $ conservation equations for the species, so the total problem involves $\numspecies+4$ equations. The stress tensor, \viscstresstensor, and the heat flux, \heatflux, are computed as 
\begin{align}
    \viscstresstensor
    &= \frac{\viscosity}{\Reynolds_\freest} \left( \grad{\velocity} + \transposep{\grad{\velocity}} - \left(\divg{\velocity}\right)\eyematrix \right),
    \label{eq:definitionstress} \\
    \heatflux
    &= - \frac{\thermconductivity}{\Reynolds_\freest \Prandtl_\freest \Eckert_\freest}
    \grad{\temperature}
       + \sumspecies \densityspecies \speciesh \diffvelocity.
    \label{eq:definitionheatflux}
\end{align}
The nondimensional Reynolds number, $\Reynolds_\freest$, and Prandtl number, $\Prandtl_\freest$, are defined in the free stream, at the domain inlet. The Eckert number, $\Eckert_\freest$, is also computed at the free stream and is equal to one by design. These nondimensional quantities are defined with dimensional quantities (distinguishable by the tilde) as 
\begin{equation}
    \Reynolds_\freest = \frac{\freestream{\density}\freestream{\soundspeed}\valref{L}}{\freestream{\viscosity}},\quad
    \Prandtl_\freest  = \frac{\freestream{\viscosity}\freestream{\heatp}}{\freestream{\thermconductivity}},\quad
    \Eckert_\freest   = \frac{\freestream{\soundspeed}^2}{\freestream{\heatp}\valref{\temperature}} \;
    \label{eq:nondimnumbers}
\end{equation}
where $\valref{\temperature} = (\freestream{\gammaratio}-1) \freestream{\temperature}$, and $\soundspeed$ stands for the speed of sound. \viscosity~ denotes the dynamic viscosity,  \thermconductivity~ the frozen thermal conductivity, and \heatp~ the specific heat at constant pressure. 

%-----------------------------------
\subsection{Thermodynamic and chemical models for a mixture in chemical non-equilibrium}
\label{sub:thermochem}

In the following, the closure of the governing equations when assuming a mixture in chemical non-equilibrium is detailed. In general, a reacting gas in a high-enthalpy flow is considered as a multi-component mixture that consists of a set of species \speciesset~ interacting through a defined network of reactions. The presence of the species makes the chemical reaction and diffusion terms in the governing equations significant, which in turn require to be modeled using various assumptions.

For a multi-component gas mixture, the global mixture properties are derived from the species properties based on
\begin{equation}
    \density = \sumspecies \densityspecies,\quad
    \internale = \sumspecies \speciesmassfrac \speciese,\quad
    \internalh = \sumspecies \speciesmassfrac \speciesh
    \label{eq:mixsums}
\end{equation}
where the species mass fraction is $\speciesmassfrac = \frac{\densityspecies}{\density}$, with $\sumspecies \speciesmassfrac = 1$. The mixture thermodynamic and transport properties depend, generally, on their composition, and any two thermodynamic properties, for example the temperature and the pressure, which define the thermodynamic state of the mixture. The composition is commonly taken as an independent variable. It can potentially become dependent on other thermodynamic quantities for the special cases of frozen chemistry or chemical equilibrium. However, these cases are not described in this study. The individual species properties are accurately computed by kinetic theory and statistical mechanics \citep{scoggins_2014_mpp, Scoggins2017}. Hence the composition and the two thermodynamic properties can be concatenated into the thermodynamic state vector $\tilde{\mathbf{q}}_{th}$. This defines the relation $\Tilde{\mathbf{Z}} = \Tilde{f}(\tilde{\mathbf{q}}_{th})$ on which the algorithm is based. \\ 

When finite-rate chemistry is not neglected, the species mass production rate, \massprod, and the species diffusion velocities, \diffvelocity, need to be modeled. 
For a general case, a set of reactions, \reactionsset, is considered depending on the mixture in question. 
Each reaction \reactionr\ is characterized by a reaction rate, \reactionrate, which is computed by the forward rate, \forwardrate, and the backward rate, \backwardrate. These rates, in turn, are obtained according to experimentally or theoretically calibrated Arrhenius formulas in the form $\forwardrate = C_\reactionr \temperature^{n_\reactionr} \exp{\left({\temperature_a}_\reactionr/\temperature\right)}$, and $\backwardrate = \forwardrate/{\equilconst{(\temperature)}}$, where \equilconst\ is the reaction equilibrium constant at the specific conditions.
This description is given in \cref{eq:reaction} for a generic reaction. 
\begin{equation}
    \text{Reaction (r):}
    \quad\quad
    \sumspecies \nu'_{\reactionr,\speciess} S_\speciess
    \xrightleftharpoons[\backwardrate]{\forwardrate} 
    \sumspecies \nu''_{\reactionr,\speciess} S_\speciess
    \quad\quad
    \label{eq:reaction}
\end{equation}
The net reaction rate is then given by
\begin{equation}
    \reactionrate = \left[ \forwardrate \Pi_\speciess \left( \frac{\densityspecies}{\mwspecies} \right)^{\nu'_{\reactionr,\speciess}}
    - \backwardrate \Pi_i \left( \frac{\densityspecies}{\mwspecies} \right)^{\nu''_{\reactionr,\speciess}} \right] 
    \cdot \sumspecies \left( Z_{\reactionr,\speciess} \frac{\densityspecies}{\mwspecies} \right)
    \label{eq:reactionrate}
\end{equation}
where the species molar mass is \mwspecies\ and the efficiency of species \speciess\ as a third-body in reaction \reactionr\ is $ Z_{\reactionr,\speciess}$.
The net mass production rates for species \speciess\ from all reactions are obtained as
\begin{equation}
    \massprod = \mwspecies \sumreactions \left( \nu''_{\reactionr,\speciess} - \nu'_{\reactionr,\speciess} \right) \reactionrate.
    \label{eq:reactionmass}
\end{equation}
The diffusion flux $\vec{J}_\speciess = \densityspecies \diffvelocity$ appearing in \cref{eq:speciesmass} also needs to be modeled. The accurate and rigorous Stefan-Maxwell multicomponent diffusion model provides the diffusion velocities $\diffvelocity$ as the solution of a constrained linear system of equations \cite{Scoggins2017}. In that framework, the local molar fraction gradients are needed, which cancels the purely local assumption of the input/output problem. Hence, in this study, we use a simple expression based on Fick's law with a mass correction \cite{hirschfelder1964,ramshaw1990self} given by 
\begin{equation}
    \vec{J}_\speciess = -c M_s D_s \grad{\speciesmassfrac} + c Y_s \sum_{\subs{i} \in \mathrm{S}} M_\subs{i} D_\subs{i} \nabla Y_\subs{i}.
    \label{eq:ramshaw}
\end{equation}
Here, $c = \sumspecies (\densityspecies / M_\speciess),$ and $D_\speciess$ is the averaged diffusion coefficient for species \speciess, defined below 
\begin{equation}
    D_\speciess = \frac{1 - \speciesmolefrac}{\sum_{\subs{r} \neq \speciess} X_\subs{r}/D_{\speciess, \subs{r}}}.
    \label{eq:speciesavgdiffusioncoef}
\end{equation}

The thermochemical library \acrshort{Mutation++} \cite{Scoggins2020} is used to compute thermodynamic and transport properties at different conditions. The reader is referred to the description provided in \cite{Scoggins2017,Scoggins2020} for further details on the library.

%-----------------------------------
\subsection{Numerical framework}

The governing equations are solved using a high-fidelity in-house flow solver for the direct numerical simulation of hypersonic flows in chemical non-equilibrium, as described in \cite{margaritis2022}. Compact schemes are used for a spatial discretization, and explicit time-integration is performed using Runge-Kutta schemes. More details about the discretization can be found in the aforementioned paper.

\subsubsection{Coupling with \acrshort{Mutation++}}

The thermodynamic and transport properties, as well as the source terms resulting from the chemical kinetics models, are extracted from the \acrshort{Mutation++} library \cite{Scoggins2020} to close the governing equations, \crefrange{eq:globalmass}{eq:totenergy}. The library, originally written in \texttt{C++}, is coupled with the numerical solver using a wrapper in \texttt{Fortran 95}. 

For the remainder of this study, we will only consider an air mixture composed of five species $\speciesset = [N_2, O_2, NO, N, O]$. However, we stress that the algorithm can be applied to any gas mixture. At each grid point, we have access to the local thermodynamic state and composition $\Tilde{\mathbf{q}}_{th}$ (in dimensional units), 
\begin{equation}
\Tilde{\mathbf{q}}_{th}=
\begin{bmatrix} 
\rho  & \rho e & \rho_N & \rho_O & \rho_{NO} & \rho_{O_2}
\end{bmatrix}.
\end{equation}
$\Tilde{\mathbf{q}}_{th}$ is then passed to \acrshort{Mutation++}, which returns the necessary thermochemical properties needed to close the reactive compressible Navier-Stokes  \crefrange{eq:globalmass}{eq:totenergy},
\begin{equation}
\Tilde{\textbf{z}}=
\begin{bmatrix} 
\mu & \kappa & P & T & h_s & \omega_s & D_s
\end{bmatrix} \quad s \in \speciesset. 
\end{equation}
Hence, calls to \acrshort{Mutation++} can be seen as an input/output problem $\Tilde{\mathbf{z}}= \Tilde{f}(\Tilde{\mathbf{q}}_{th})$ where the function $\Tilde{f}$ represents the library. 

%-----------------------------------
\subsection{Case 1: Mach-10 adiabatic boundary layer}\label{simu}
In this section, the first simple test-case used to showcase the applicability and performance of the algorithm is presented. We simulate a Mach-10 adiabatic flat-plate boundary layer in Earth atmosphere, based on the work of Marxen \textit{et al}  \citep{Marxen2013,Marxen2014}. The thermodynamic and freestream conditions are presented in \cref{table:condition_olaf}. Details of the numerical grid can be found in \cite{margaritis2022}.
\begin{table}[htbp]
\begin{center}
\begin{tabular}{m{0.25\textwidth}c}
 \multicolumn{2}{c}{Test case 1} \\
 \hline
  \hline
 $ M_\infty $ & 10\\ 
 $ Re_\infty $ & $10^5$\\ 
 $ T_\infty [K] $ & 350  \\ 
 $ p_\infty [Pa] $ & 3596 \\ 
 \hline
 \hline
\end{tabular}
\caption{Thermodynamic and free stream conditions for Mach-10 adiabatic flat-plate boundary layer test case (from \cite{Marxen2013}).}
\label{table:condition_olaf}
\end{center}
\end{table} 

\begin{figure}[htbp]
    \centering

    \subfloat[\label{subfig:vel_A}]{%
      \includegraphics[width=0.3\columnwidth]{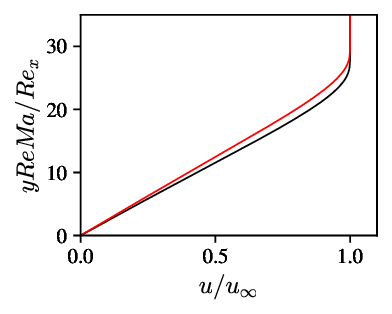}
    }
    \subfloat[\label{subfig:temp_A}]{%
      \includegraphics[width=0.3\columnwidth]{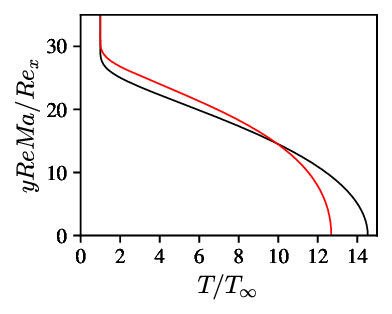}
    }
    \subfloat[\label{subfig:species_A}]{%
      \includegraphics[width=0.3\columnwidth]{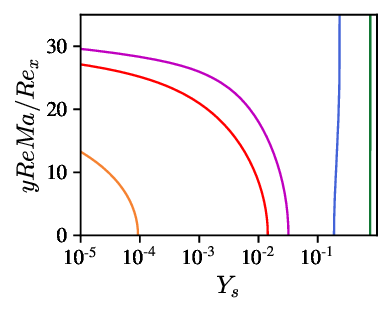}
    }
    \caption{Profiles of (a) streamwise velocity, (b) temperature, (c) species mass fractions from left to right $N$, $NO$, $O$, $O_2$ and $N_2$, at $Re_x = 2000$. (a,b) Red line: simulation in chemical non-equilibrium, black line: perfect gas assumption. }\label{fig:profiles_caseA}
\end{figure}

This case has already been validated against literature with the current flow solver in \cite{margaritis2022}. Due to the wall temperature approaching $T_{wall}\approx 4900K$ near the inflow, $N_2$ and $O_2$ rapidly start to produce their monoatomic counterpart, as well as $NO$, through endothermic chemical reactions. This is illustrated in \cref{subfig:species_A} presenting all the mass fraction profiles at the streamwise location of $Re_x = 2000$. Close to the wall, the $O_2$ mass fraction decreases while $O$ and $NO$ are created. To a smaller extent, $N$ is also created through $N_2$ dissociation. The mean flow and temperature profiles at a streamwise Reynolds number of $Re_x = 2000$ are presented in \cref{subfig:vel_A} and \cref{subfig:temp_A}, alongside the same case simulated using a thermally perfect gas assumption (ignoring non-equilibrium effects). The base flows differ significantly depending on the assumption used for the gas. Wall temperature decreases significantly from a thermally perfect gas to a finite-rate chemistry assumption, and the boundary layer becomes thicker. Overall, these results highlight and corroborate the importance of including chemical non-equilibrium effects for the accurate simulation of high-enthalpy hypersonic flows. On another note, it has been shown in \cite{margaritis2022} that the choice of the diffusion model (i.e Stefan-Maxwell or an approximation based on Fick's law (\cref{eq:ramshaw})) does not have any noticeable impact on the baseflow obtained in a chemical non-equilibrium simulation. However, using the algebraic equation reduces the CPU cost of the simulation by 40\% compared to the rigorous Stefan-Maxwell model.

\subsection{Case 2: Mach-5.92 shock-wave boundary layer interaction}\label{sbli}
\begin{figure}[htbp]
    \centering

    \subfloat[\label{subfig:cf_sbli}]{%
      \includegraphics[width=0.3\columnwidth]{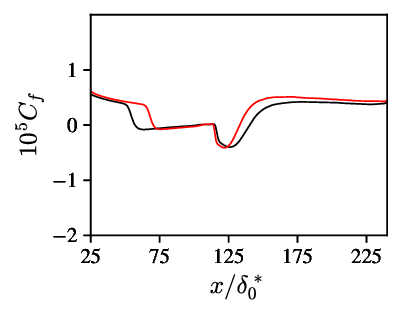}
    }
    \subfloat[\label{subfig:pres_sbli}]{%
      \includegraphics[width=0.285\columnwidth]{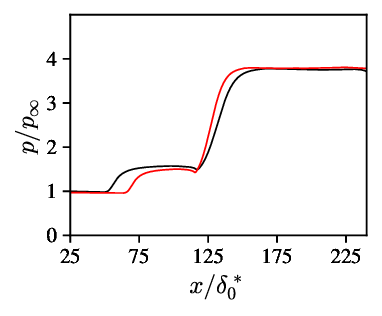}
    }
    \subfloat[\label{subfig:species_sbli}]{%
      \includegraphics[width=0.3\columnwidth]{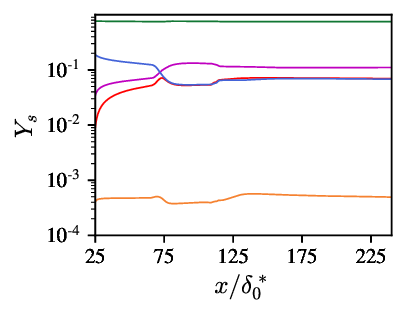}
    }
    \caption{Wall distribution of (a) skin-friction coefficient, (b) pressure, and (c) species mass fractions (from top to bottom on left side: $N_2$, $O2$, $O$, $NO$ and $N$). (a,b) Red line: simulation in chemical non-equilibrium, black line: perfect gas assumption. }\label{fig:profiles_sbli}
\end{figure}

While the adiabatic boundary layer is in a chemical non-equilibrium regime, it lacks typical compressible flow features such as shocks and expansion fans that could challenge the performance of the algorithm. To verify the applicability of the developed technique, a second case, namely a shock wave boundary layer interaction (SBLI), is proposed. This case has been initially studied by Margaritis \textit{et al.} \cite{margaritis2022}. The thermodynamic and freestream conditions are presented in \cref{table:condition_sbli} where $x_0$ denotes the shock impinging location and $\theta$ denotes the shock angle. The wall is considered as adiabatic. Details of the numerical grid and the computational setup can be found in \cite{margaritis2022}. \cref{subfig:cf_sbli} and \cref{subfig:pres_sbli} present the wall distribution of skin-friction coefficient and wall pressure, respectively. The recirculation bubble is smaller in the case in chemical non-equilibrium compared to the case assuming perfect gas. This is explained by the high concentration of atomic species in the recirculation bubble, as seen in \cref{subfig:species_sbli}. The endothermic reactions extract energy out of the bubble. Moreover, the skin-friction and pressure are higher just upstream of the reattachment point compared to the perfect-gas simulation. Hence, this case also highlights the effect of chemical non-equilibrium on the resulting flow features.

\begin{table}[htbp]
\begin{center}
\begin{tabular}{m{0.25\textwidth}c}
 \multicolumn{2}{c}{Test case 2} \\
 \hline
  \hline
 $ M_\infty $ & 5.92\\ 
 $ Re_{x_0} $ & $1.15 \times 10^6$\\ 
 $ T_\infty ~~[K] $ & $1110.5$  \\ 
 $ p_\infty ~~[Pa] $ & $61 \times 10 ^3$ \\ 
 $ \theta ~~~~[deg] $ & 13.0 \\ 
 \hline
 \hline
\end{tabular}
\caption{Thermodynamic and free stream conditions for Mach-5.92 shock-wave boundary layer case (from \cite{margaritis2022}).}
\label{table:condition_sbli}
\end{center}
\end{table} 

%---------------------------------------------------------------

\section{Results}\label{results}

The results are presented in three parts. First, each step of the off-line training is assessed, and then the performance of the model in predicting the output quantities when coupled to the Navier-Stokes solver is analyzed on the Mach 10 adiabatic boundary layer. Finally, the same strategy is applied to a shock-wave boundary layer interaction scenario.

\subsection{Model training}\label{application}
$N = 100,000$ thermodynamic state vectors $\tilde{\textbf{q}}_{th}$ are sampled from the converged solution and concatenated into the input dataset $\Tilde{\mathbf{X}}$. The corresponding \acrshort{Mutation++} output dataset $\Tilde{\mathbf{Z}}$ is also obtained. 

\subsubsection{Input/output encoding}
\label{IOE-training}
The architecture and hyperparameters of the input/output-encoder used for this test case are provided in \cref{table:architecture_IOE}. 

\begin{figure}[htbp]
    \centering
    \includegraphics[width=0.7\textwidth]{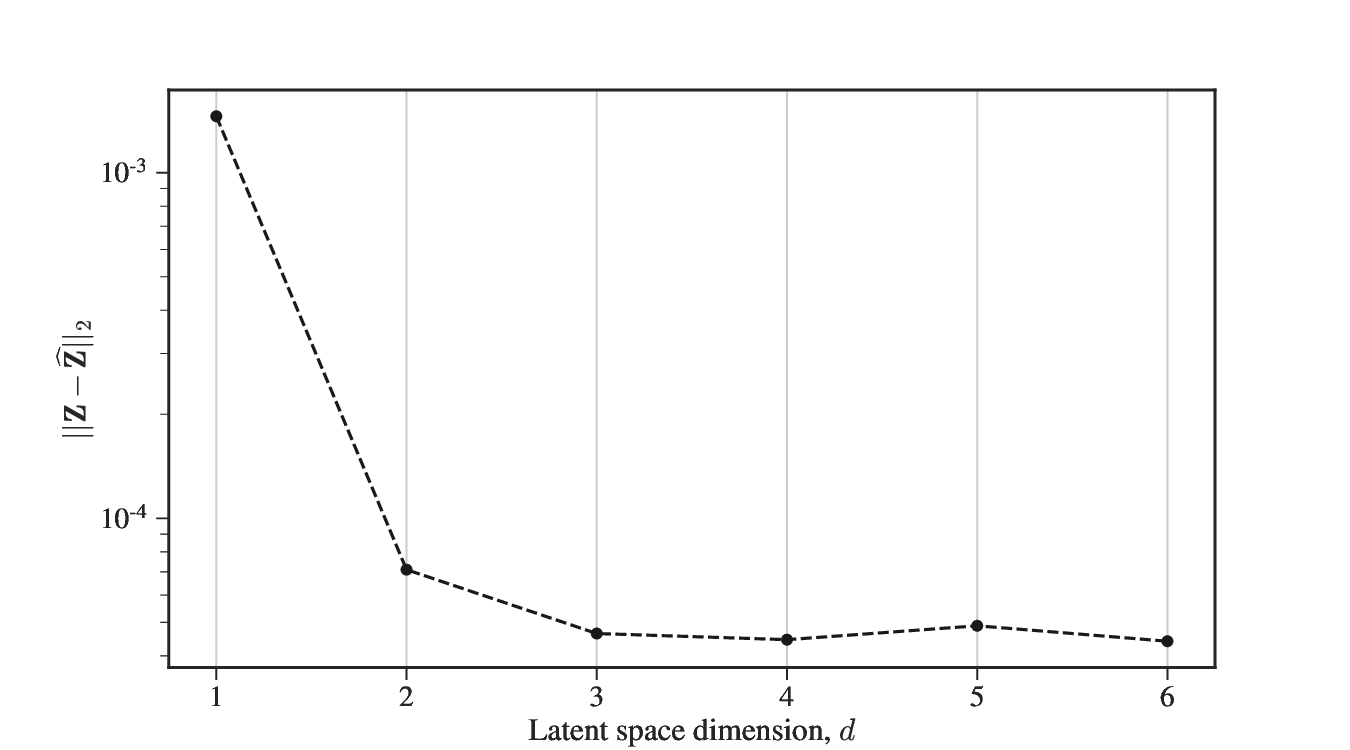}
    \caption{Reconstruction loss $\lVert  \widehat{\mathbf{Z}} -\mathbf{Z} \rVert_2$ of the embedding done by IO-E with respect to the number of latent space dimensions $d$.}
    \label{fig:elbow}
\end{figure}

\begin{table}[htbp]
\begin{center}
\begin{tabular}{ m{0.2\textwidth} m{0.1\textwidth} | m{0.2\textwidth} m{0.1\textwidth} } 
 \multicolumn{4}{c}{\textbf{Architecture}} \\
 \hline
 \hline
  \multicolumn{2}{c}{\textbf{Encoder}} & \multicolumn{2}{c}{\textbf{Decoder}}\\
  \multicolumn{1}{m{0.2\textwidth}}{Layer} & \multicolumn{1}{m{0.1\textwidth}}{Size} &  \multicolumn{1}{m{0.2\textwidth}}{Layer} & \multicolumn{1}{m{0.1\textwidth}}{Size} \\
  \hline 
  Input  & 6 & Latent space & 2 \\ 
  Fully connected & 12 & Fully connected & 6 \\ 
  Fully connected & 6 & Fully connected & 12 \\ 
  Latent space & 2 & Output & 18 \\ 
  \hline 
  \hline 
\end{tabular}
\end{center}
\vspace{3mm}
\begin{center}
\begin{tabular}{ m{0.2\textwidth} m{0.2\textwidth} | m{0.2\textwidth} m{0.2\textwidth} }
 \multicolumn{4}{c}{\textbf{Hyperparameters}} \\
 \hline
 \hline
  \multicolumn{1}{m{0.2\textwidth}}{Parameter} & \multicolumn{1}{m{0.2\textwidth}}{Value} &  \multicolumn{1}{m{0.2\textwidth}}{Parameter} & \multicolumn{1}{m{0.2\textwidth}}{Value} \\
  \hline 
  Learning rate & $1\times 10^{-3}$ & Epochs & 2000 \\ 
  Loss & Mean-squared error & Batch size & 256 \\ 
  Activation function & tanh & Optimizer & Adam (keras default) \\ 
  \hline 
  \hline
\end{tabular}
\caption{Architecture and hyperparameters of the input/output-encoder used to train the IO-E network on the $Ma=10$ adiabatic flat plat boundary layer test case.}
\label{table:architecture_IOE}
\end{center}
\end{table} 
An important parameter is the number of dimensions required to properly unfold the input manifold. To this end, the architecture and hyperparameters, as well as the random seed to initialize the network weights, are held constant, and only the dimension of the latent space is varied. We follow the reconstruction loss $\lVert  \widehat{\mathbf{Z}} -\mathbf{Z} \rVert_2 $ of the testing set as a function of the latent space dimension $d$. As seen in \cref{fig:elbow}, the loss saturates after $d=2$, suggesting that two dimensions are sufficient to represent the input manifold, originally in $\mathbb{R}^6$.

\subsubsection{Newman's community clustering}
\label{cluster-training}
\begin{figure}[htbp]
    \centering
    \includegraphics[width=0.7\textwidth]{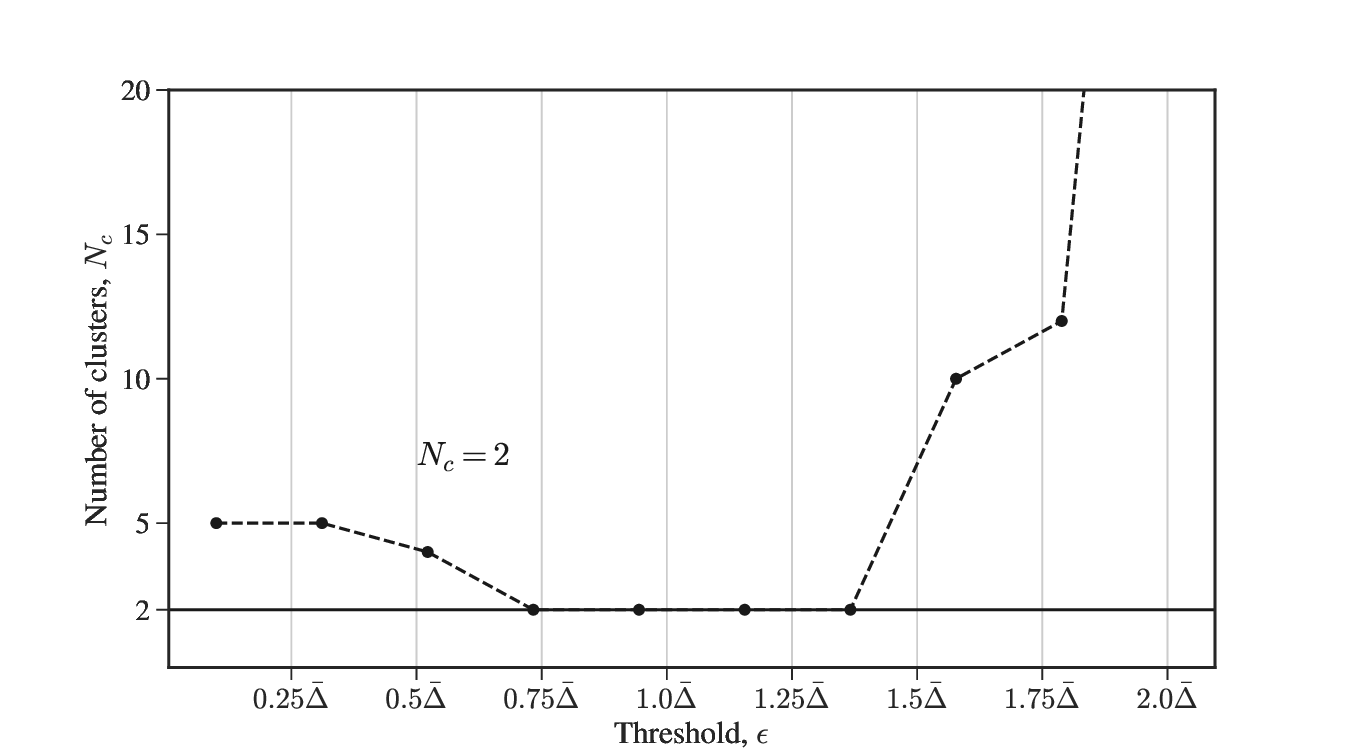}
    \caption{Number of clusters $N_c$ obtained with Newman's algorithm as a function of the threshold $\epsilon$, given as multiples of the distance matrix mean $\overline{\Delta}$.}
    \label{fig:nci_vs_eps}
\end{figure}

The second step is to cluster the data on the low-dimensional manifold using Newman's algorithm. The only hyper-parameter needed for Newman's clustering algorithm is the threshold $\epsilon$ to determine the adjacency matrix $A$ from the distance matrix $\Delta$, as described in \cref{clustering}. To showcase the robustness of the number of clusters with respect to the threshold $\epsilon$, the algorithm is applied over a range of thresholds, chosen as multiples of the mean of the distance matrix $\overline{\Delta}$. As shown in \cref{fig:nci_vs_eps}, over the range tested, the algorithm returns $N_c=2$ clusters before over-fitting with extreme values. \\

\begin{figure}[htbp]
    \centering

    \subfloat[\label{subfig:clusters_embedded}]{%
      \includegraphics[width=0.4\textwidth]{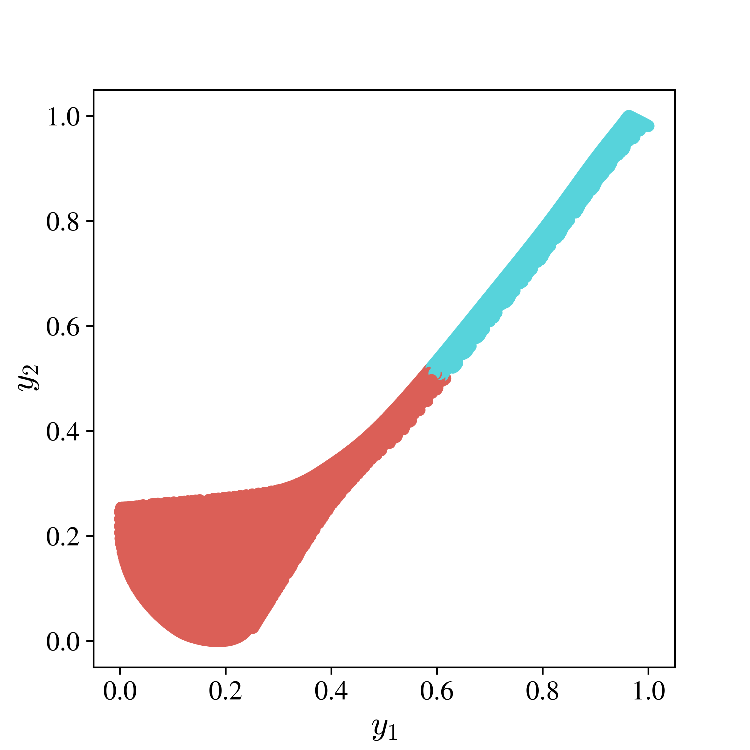}
    }
    \subfloat[\label{subfig:clusters_flow}]{%
      \includegraphics[width=0.55\textwidth]{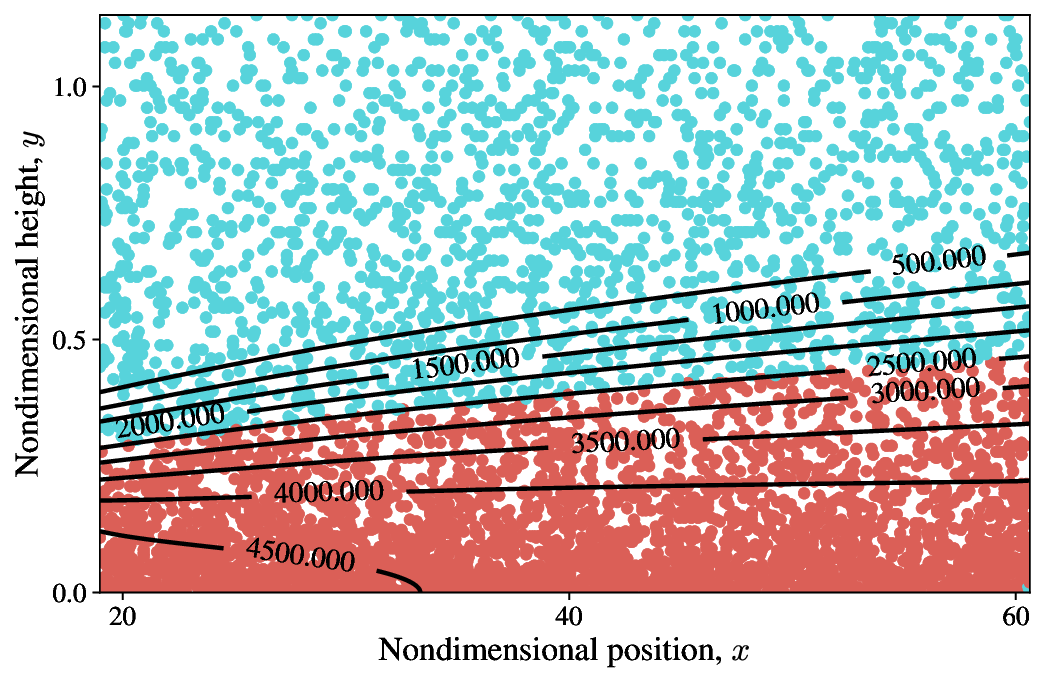}
    }

    \caption{Training points $\mathbf{Y}$ colored by their cluster number. (a) In the latent space found by IO-E. (b) At the Cartesian location they were sampled from, with contours of temperature $T$ in Kelvin.}\label{fig:clusters_caseA}
\end{figure}

\cref{subfig:clusters_embedded} shows the two clusters obtained in the embedded space. As expected, the two clusters define different regions in the reduced space. To gain more physical insight,
\cref{subfig:clusters_flow} shows randomly selected points of the training set, mapped back to their original location in the Cartesian space and colored by their cluster number. Contours of temperature $T$ in Kelvin are added. Each cluster fills a region of the flow with different levels of chemical non-equilibrium. The blue cluster represent the free stream, where temperature is low and chemistry is frozen. Alternatively, the red cluster corresponds to the near-wall region with dissociated species and high temperatures. 

The random forest classifier has been trained with $n_{tree} = 20$. This number proved sufficient to obtain an acceptable prediction accuracy of new points clusters, as seen in \cref{fig:confusion matrix} above. 

% \begin{figure}[htbp]
%     \centering
%     \begin{subfigure}[b]{0.4\textwidth} 
%         \centering \includegraphics[width=\textwidth]{Figures/Clusters_reduced_space.eps}
%         \label{fig:clusters_embedded}
%     \end{subfigure}
%     ~ % ce symbole ajoute un espacement horisontal entre les premières deux images
%     \begin{subfigure}[b]{0.55\textwidth}
%         \centering \includegraphics[width=\textwidth]{Figures/clusters_in_flow.eps}
%         \label{fig:clusters_flow}
%     \end{subfigure}
%     \caption{Training points $\mathbf{Y}$ colored by their cluster number. (a) In the latent space found by IO-E. (b) At the Cartesian location they were sampled from, with contours of temperature $T$ in Kelvin.}\label{fig:clusters_caseA}
% \end{figure}

\subsubsection{Surrogate model construction}
\label{rbf-training}
\begin{figure}[htbp]
    \centering
    \includegraphics[width=0.7\textwidth]{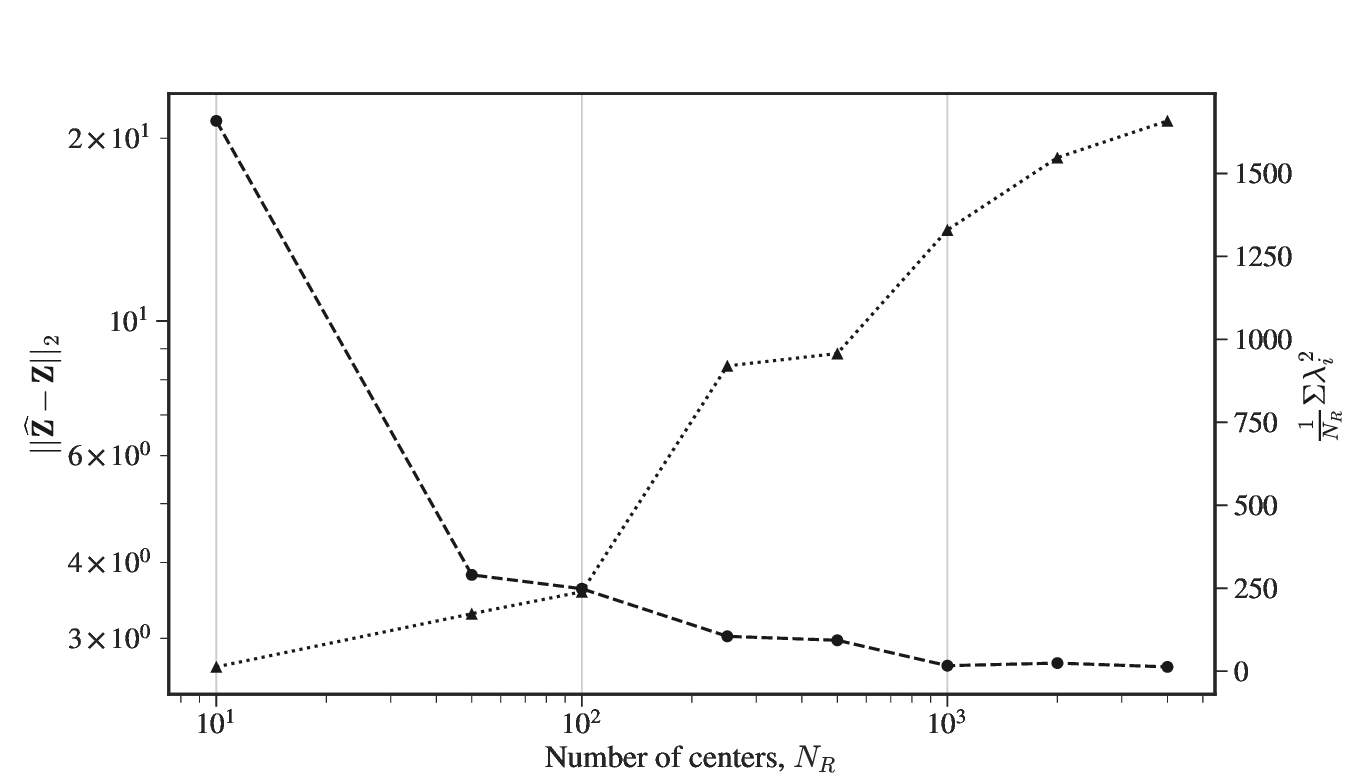}
    \caption{Left axis: reconstruction error of surrogate model $\lVert  \widehat{\mathbf{Z}} -\mathbf{Z} \rVert_2$ on a testing set (dashed line). Right axis: mean of squared RBF coefficients $\overline{\Lambda^2}$ (dotted line). Both with respect to the number of RBF centers $N_R$ used in training.}
    \label{fig:loss_centroids}
\end{figure}

Once the clusters are determined, a separate RBF interpolant is trained for each cluster as described in \cref{MR}. To enforce continuity of the surrogate through the clusters, the nearest centroids shared between the two clusters are added to the training set. \\
To assess the number of centroids needed, the RBFs are trained simultaneously, for both clusters, with the same number of centroids $N_R$ (note that this number can be varied to accommodate clusters of different sizes). The error of the model for the testing set $\lVert  \widehat{\mathbf{Z}} -\mathbf{Z} \rVert_2$ is  plotted against the number of centroids $N_R$ in \cref{fig:loss_centroids} (blue curve). As expected, the error decreases as the number of centroids is increased. To assess potential overfitting, one can follow the evolution of the mean of the squared RBF coefficients $\overline{\Lambda^2}$,
\begin{equation}
    \overline{\Lambda^2} = \frac{1}{N_R}\sum_{i=1}^{N_R} \lambda_i^2
\end{equation}
where a high value will likely indicate overfitting. On the right axis of \cref{fig:loss_centroids}, we see that $\overline{\Lambda^2}$ increases with the number of RBF centroids. Therefore, choosing the right $N_R$ is a trade-off between low error and low overfitting. In this case, a value of $N_R=250$ for each cluster has been retained.

\subsection{Model accuracy}\label{mode_acc}
The reduced library is tested (off-line) on a full snapshot (which includes also the training points used to build the model) to assess the capacity of the model to interpolate new points not encountered during training. Four configurations of the data-driven model are tested: (i) model 1, using the full IO-E for prediction (ii) model 2, with no dimensionality reduction and no clustering ($d=6$, $Nc=1$, $N_R = 250$), (iii) model 3, with dimensionality reduction, but without clustering, ($d=2$, $Nc=1$,$N_R = 250$), and (iv) model 4,  with both dimensionality reduction and clustering ($d=2$, $Nc=2$,$N_R = 250$).

\begin{figure}
    \centering

    \subfloat[\label{subfig:model0}]{%
      \includegraphics[width=0.45\columnwidth]{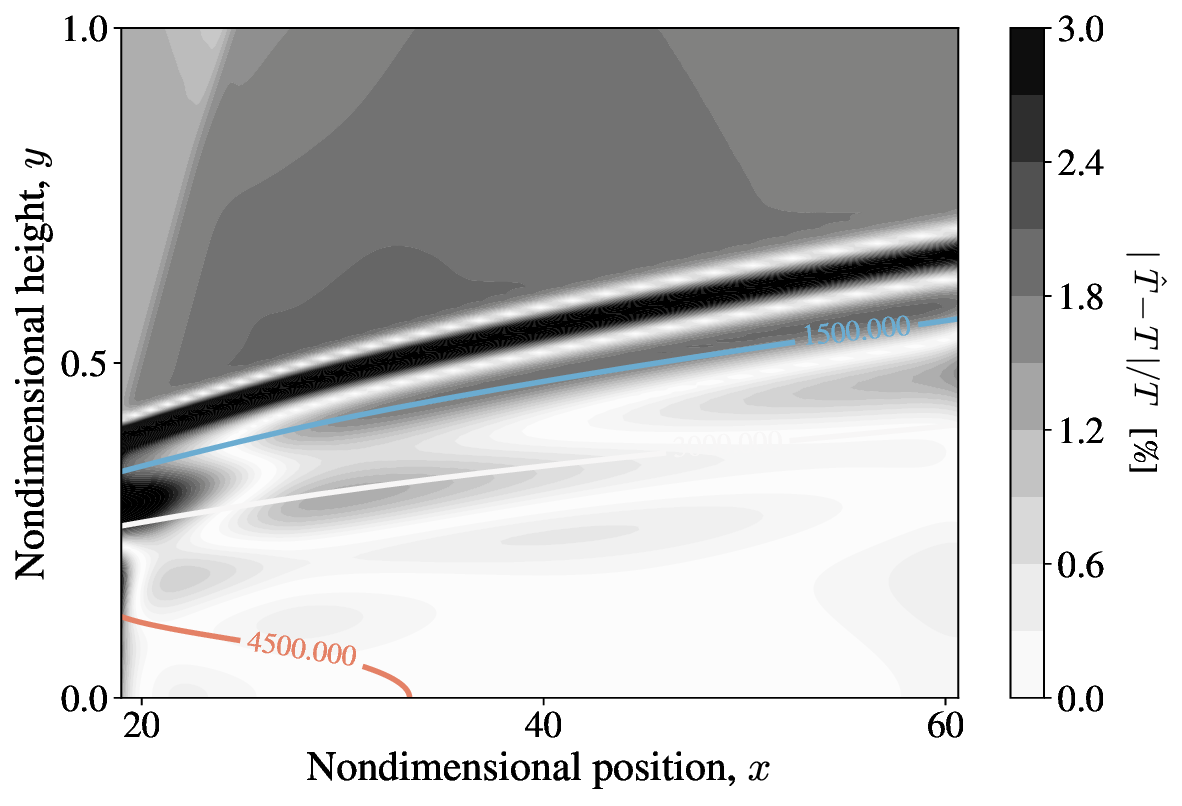}
    }
    \subfloat[\label{subfig:model1}]{%
      \includegraphics[width=0.45\columnwidth]{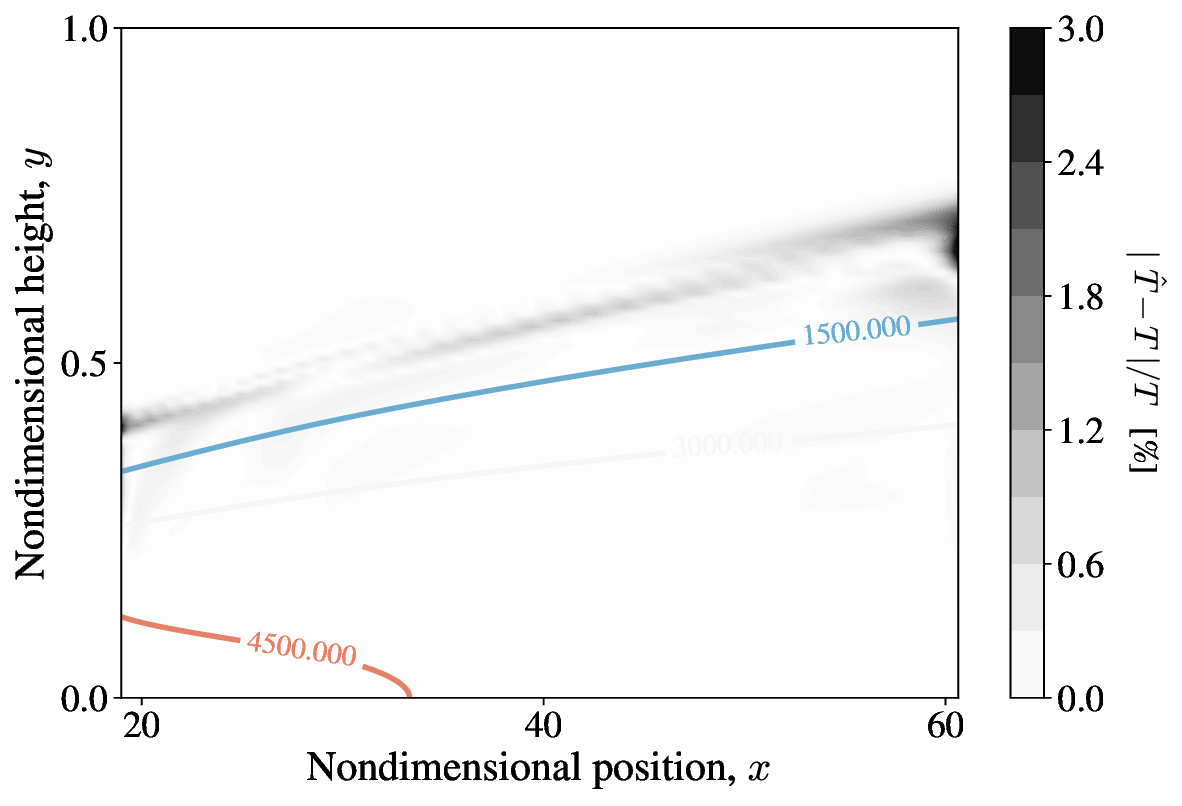}
    }
    
    \subfloat[\label{subfig:model2}]{%
      \includegraphics[width=0.45\columnwidth]{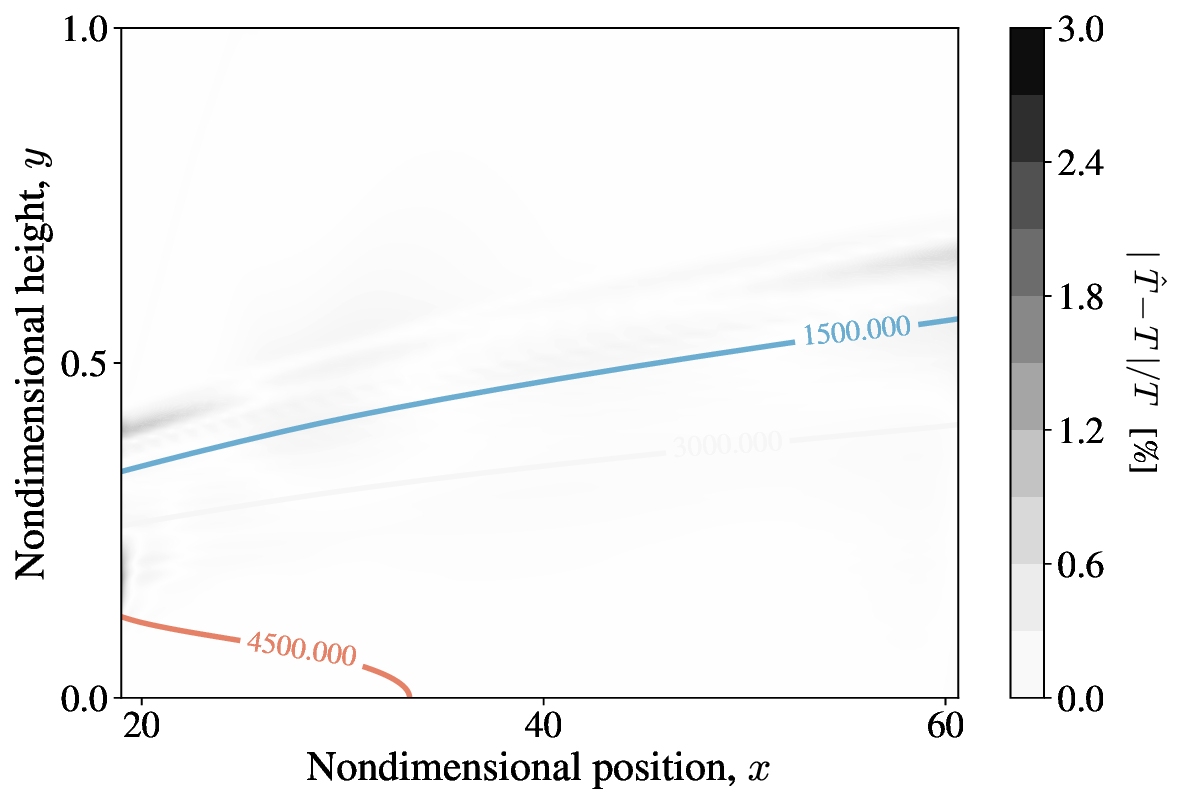}
    }
    \subfloat[\label{subfig:model3}]{%
      \includegraphics[width=0.45\columnwidth]{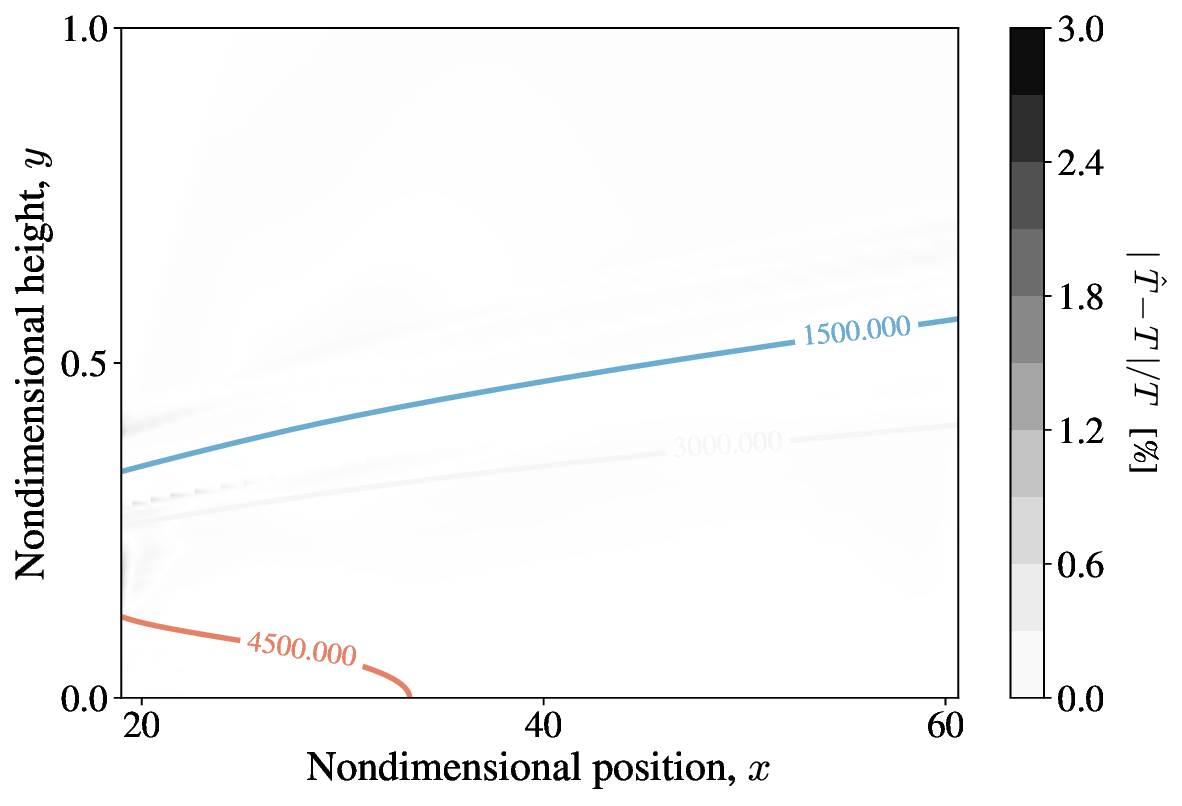}%
    }
    \caption{Comparison of the relative temperature error $\abs{\hat{T}-T}/T$ in percent with contours of temperature. (a) Model 1: full IO-E (b) Model 2: $d=6, N_c=1, N_R=250$ (c) Model 3: $d=2, N_c=1, N_R=250$ (d) Model 4: $d=6, N_c=2, N_R=250$}\label{fig:full_scale}
\end{figure}

\cref{fig:full_scale} displays the relative error (as a percentage) between the temperature $T$ given by \acrshort{Mutation++} and the prediction of the data-driven model $\hat{T}$, for the four configurations enumerated above; contours of temperature in Kelvin are also added to highlight the evolution of the flow. First, the prediction of model 1 produces the highest error by a wide margin. This showcases the difficulty of properly training a neural network for prediction in high dimensions. The figure shows the maximum relative error in all three remaining models to be only a few percent, and located around the edge of the boundary layer, where the gradients are strongest. In addition, the error in model 2 is higher than the error of model 3, even though some information is lost in the latter due to the encoding step. This is a direct consequence of the curse of dimensionality: as the number of dimensions increases, the sampling volume in the input space increases exponentially. However, in this case, we kept the number of RBF centers $N_R$ fixed. To get to the same level of accuracy, $N_R$ should be increased in model 2, resulting in a performance loss. Finally, the error decreases slightly from model 3 to model 4. This improvement originates from the clustering step. In fact, each of the two clusters have $N_R=250$ centers. Hence, the input space is actually populated with $N_R = 500$ RBF centers in model 4. According to \cref{fig:loss_centroids}, this likely improves the accuracy of the surrogate surface. The added number of centroids, however, does not result in a loss of performance. Assuming that a fraction $\alpha$ of the $N_t$ query points are in cluster 1, then $1-\alpha$ query points are in cluster 2. The evaluation time of the surrogate surface linked to cluster 1 then scales roughly as $\alpha d N_t N_R C_{RBF}$. Similarly, for cluster 2, it scales as $(1-\alpha) d N_t N_R C_{RBF}$. Finally, the total evaluation time, i.e., the sum of the two, remains $d N_t N_R C_{RBF}$. This can be easily generalized to a higher number of clusters. \\
These results demonstrate that the preprocessing steps involved in the construction of the model improve overall performance while maintaining a high level of accuracy.

\subsection{Model stability}
\begin{figure}[htbp]
    \centering

    \subfloat[\label{subfig:vel_ML}]{%
      \includegraphics[width=0.3\columnwidth]{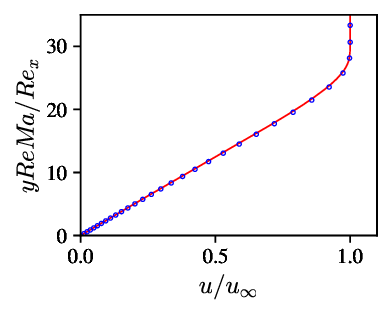}
    }
    \subfloat[\label{subfig:temp_ML}]{%
      \includegraphics[width=0.3\columnwidth]{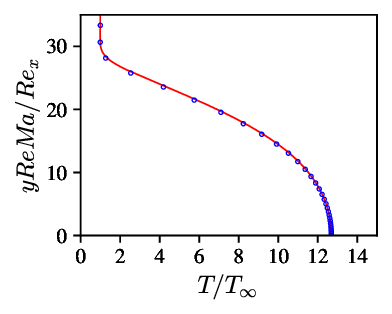}
    }
    \subfloat[\label{subfig:species_ML}]{%
      \includegraphics[width=0.3\columnwidth]{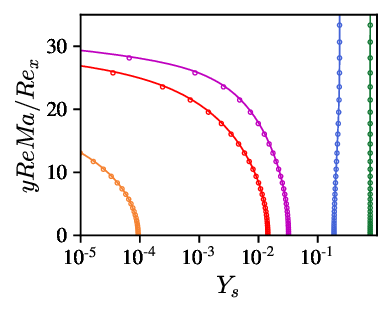}
    }
    \caption{Comparison of profiles of (a) streamwise velocity, (b) temperature, (c) species mass fractions from left to to right $N$, $NO$, $O$, $O_2$ and $N_2$ at $Re_x = 2000$. Solid line and symbols correspond to the solution obtained using \acrshort{Mutation++} and the data-driven model, respectively.}\label{fig:baseflow_mppml}
\end{figure}

The resulting data-driven model (model 4, with all pre-processing steps) is coupled to the flow solver in a time-marching simulation. Starting from the solution obtained with \acrshort{Mutation++}, the simulation is restarted using the reduced library only, also referred as a closed-loop prediction. After running for a couple of flow-through times, the solution remains stable. The base-flow profiles are compared for various quantities of interest in \cref{fig:baseflow_mppml}. Excellent agreement is found between the profiles. This validates the accuracy and suitability of the data-driven model to simulate hypersonic flows in chemical non-equilibrium over the enthalpy range observed during the training step.

\subsection{Model performance}
\begin{figure}[htbp]
    \centering
    \includegraphics[width=0.55\textwidth]{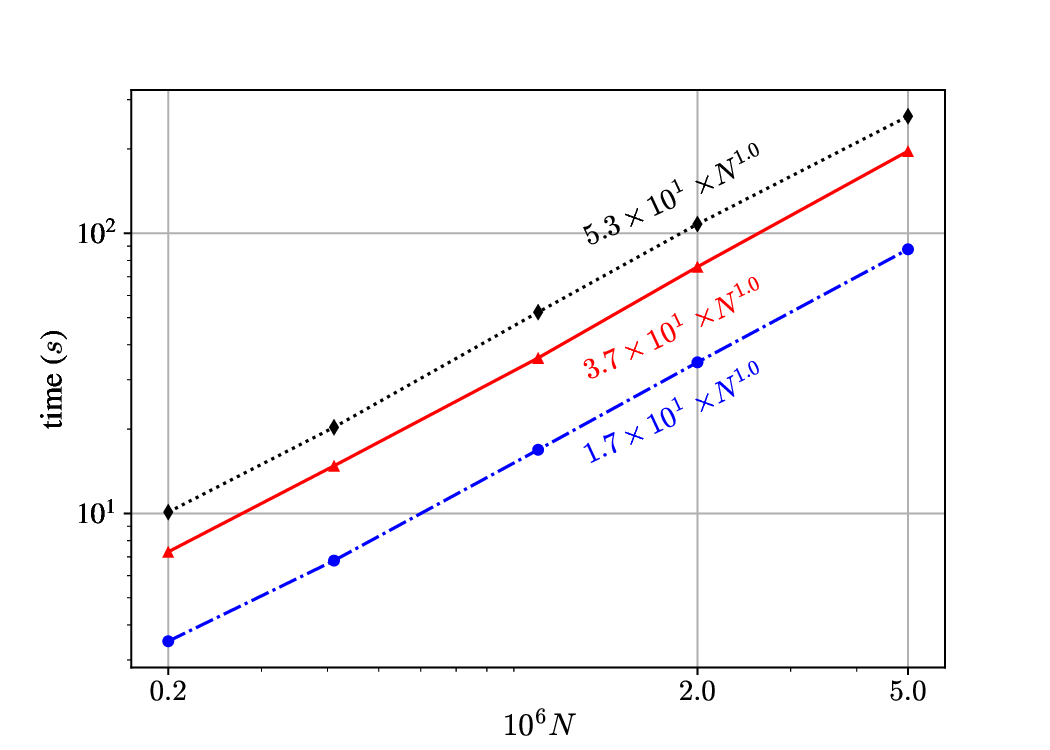}
    \caption{Comparison of the time complexity of \acrshort{Mutation++} with Stefan-Maxwell diffusion (dotted black line), the purely local version (solid red line) and the data-driven model (dash dotted blue line). The best non-linear least-squares fit of the form $CN^{\alpha}$ is added.}
    \label{fig:time_complex}
\end{figure}

To compare the performance of the data-driven model to the full library, we performed a scaling study. 
\acrshort{Mutation++} is a serial library, hence its time complexity can be expressed as $O(C_{M++}N_t)$ where $N_t$ is the number of independent, evaluated thermodynamic states. Two variants of \acrshort{Mutation++} are considered here. The first one solves directly the Stefan-Maxwell diffusion problem and returns the diffusion velocity. In the second one, the diffusion coefficients $D_\speciess$ are returned and the diffusion fluxes are later computed using \cref{eq:ramshaw}. The diffusion fluxes are computed in the same fashion with the data-driven model. Moreover, we recall that its time complexity is $O(C_{ML}N_t)$, where $C_{ML} = O(HC_{ac}L + n_{tree}depth + N_R C_{RBF})$. For all three thermochemical models, the prefactor is empirically determined in \cref{fig:time_complex}.

All curve fits, shown in \cref{fig:time_complex}, suggest that, in practice, the models scale as $O(N_t)$ with exponents close to unity. The ratio of the prefactor is $C_{ML}/C_{M++} \approx 0.45$ when using \acrshort{Mutation++} in its local version. We can therefore expect a 55\% CPU gain by using the data-driven model instead of \acrshort{Mutation++}. In fact, we assessed a CPU time reduction of 50\%  during the simulation, with a grid of size $N \approx 400,000$. 
This confirms a speedup through the use of a surrogate model. Secondly, the speed-up is even more significant when the library also solves for the Stefan-Maxwell diffusion problem at each grid point. Although not rigorously a one-to-one model comparison anymore, the data-driven model now performs 70\% faster without any loss of accuracy. In fact, Fick's law based diffusion model have been shown to be highly accurate in hypersonic simulations.
Moreover, fine tuning of the hyperparameters may allow even higher CPU gains as $C_{ML}$ is proportional to a linear combination of the hyperparameters. Finally, we stress that the data-driven algorithm is a python implementation competing with a compiled C++ library. The speed-up reported here can be significantly increased by porting the model to a compiled language. It is also believed that the speed-up would be more significant when the dimensionality of the input space increases to include more chemical species.
We thus expect even larger CPU gains in the future with an optimized implementation and added adaptivity.

\subsection{Application to the SBLI case}
\begin{figure}[htbp]
    \centering

    \subfloat[\label{subfig:clusters_embedded_sbli}]{%
      \includegraphics[width=0.3425\textwidth]{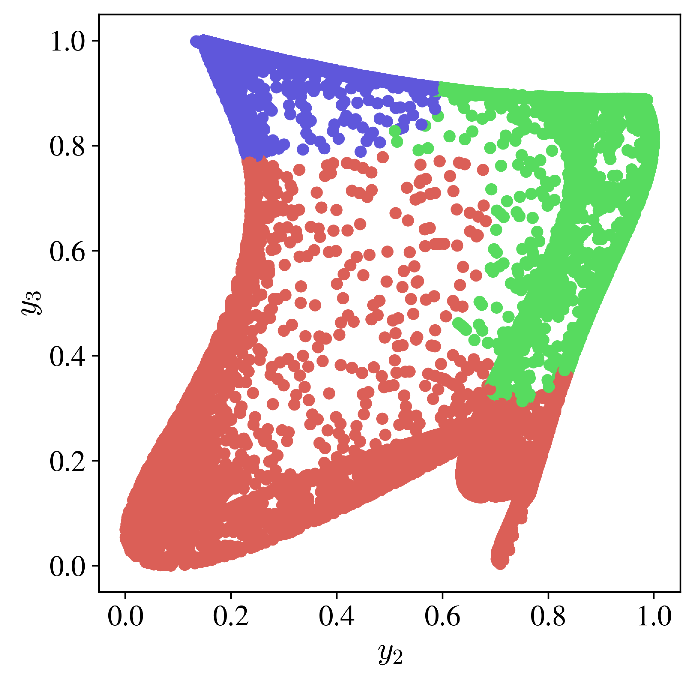}
    }
    \subfloat[\label{subfig:clusters_flow_sbli}]{%
      \includegraphics[width=0.6\textwidth]{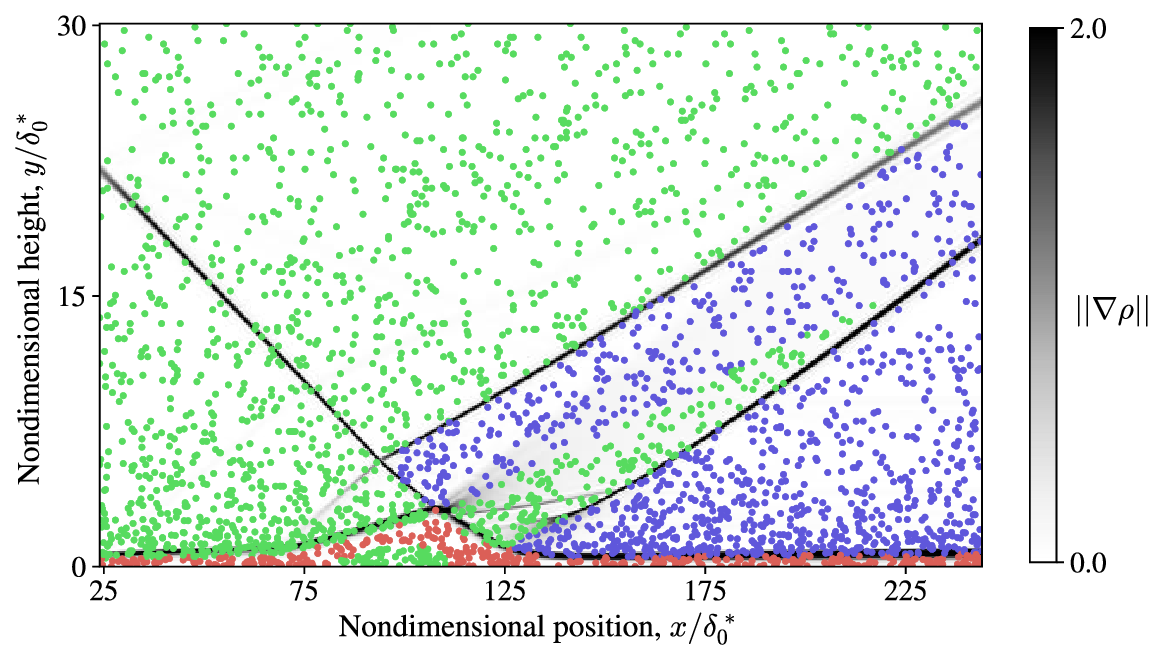}
    }

    \caption{Training points $\mathbf{Y}$ colored by their cluster number. (a) In the latent space found by IO-E. (b) At the Cartesian location they were sampled from, with contours of magnitude of the density gradient $\lVert \nabla \rho \rVert$ }\label{fig:clusters_sbli}
\end{figure}

Following the same steps, the model is trained on the SBLI case. The model has the following specifications: $d=3$, $c=3$, $N_R=250$. In this case, the dimensions of the latent space and the number of clusters are higher due to the more complex thermodynamic manifold learned by the IO-E. A two dimensional projection of the three dimensional manifold is presented in \cref{subfig:clusters_embedded_sbli}. In this plane, distinct thermodynamic regions (i.e. different clusters) are wrapped around a scarcely populated center area. This can be explained by the impinging, recirculation and reflected shocks that induce abrupt change in the thermodynamic state.
These regions and their borders become even more meaningful when reported to their physical location in the flow, as seen on \cref{subfig:clusters_flow_sbli} where a numerical Schlieren is superposed. The green cluster corresponds to mildly hot conditions with high density, i.e. the freestream and post impinging shock conditions. After the recirculation shock, the thermodynamic states shift instantaneously to higher densities and temperatures, represented by the blue cluster. However, close to the apex of the recirculation bubble, the expansion fan decreases these thermodynamic variables, inducing a shift back to the green cluster. Finally, the red cluster, found in the boundary layer, corresponds to high temperatures and low densities. At the core of the recirculation bubble, temperature decreases and density increases, which brings the local state vector back to the green cluster.

\begin{figure}[htbp]
    \centering

    \subfloat[\label{subfig:Cf_ML}]{%
      \includegraphics[width=0.3\columnwidth]{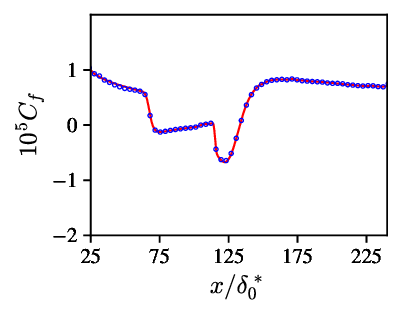}
    }
    \subfloat[\label{subfig:pres_ML}]{%
      \includegraphics[width=0.285\columnwidth]{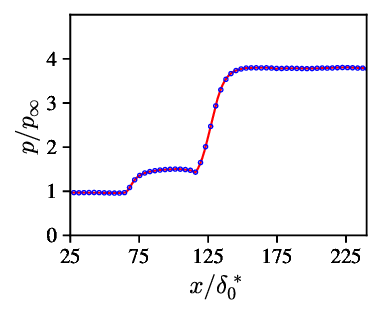}
    }
    \subfloat[\label{subfig:species_sbli_ML}]{%
      \includegraphics[width=0.3\columnwidth]{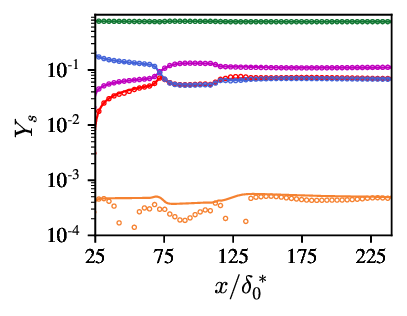}
    }
    \caption{Comparison of wall-distribution of (a) skin-friction coefficient, (b) pressure, and (c) species mass fractions (from top to bottom on left side: $N_2$, $O2$, $O$, $NO$ and $N$). Solid line and symbols correspond to the solution obtained using \acrshort{Mutation++} and the data-driven model, respectively.}\label{fig:sbli_mppml}
\end{figure}

In a closed-loop simulation, the model remains stable while maintaining a factor 2 speed-up in predicting thermochemical properties. In fact, \cref{subfig:pres_ML} and \cref{subfig:Cf_ML} show that the wall pressure and skin friction remain in excellent agreement with the baseflow solution after  2 flow-through times. The only discrepancy with the initial solution is observed for atomic nitrogen mass concentration in \cref{subfig:species_sbli_ML}. However, it is present in such small quantities that it does not have an impact on the stability of the solution. In fact, the thermochemical properties are not sensitive to small perturbations in atomic nitrogen concentration (cf. \cref{fig:mpp_IO}), another fact motivating the dimensionality reduction performed in pre-processing.  

%---------------------------------------------------------------
\section{Conclusions}\label{sec4}
In this paper, we presented a novel technique to reduce any high-dimensional look-up library to a lower-dimensional surrogate, and thus reduce the CPU costs of numerical simulations that rely on these libraries. Several machine learning techniques have been used: encoding based on deep neural networks, community clustering, surrogate modeling and classification in a three-step learning phase. In the first step, the proposed input/output-encoder architecture has been shown to outperform partial least-squares (PLS) for dimensionality reduction of input/output relations. Clustering was performed using Newman's algorithm. It discovered physically consistent clusters in the low-dimensional latent space without \textit{a-priori} knowledge of the number of clusters. Then, a random-forest classifier was trained, which reliably predicted the cluster of previously unencountered data points. Finally, a radial basis function network was constructed on each cluster to obtain a continuous and local representation of the library via a reduced-order surrogate model. The combination of these pre-processing steps has been shown to improve the efficiency of the model on our two test cases: a Mach-10 adiabatic boundary layer and a Mach-5.92 shock wave boundary layer interaction, both in chemical non-equilibrium. After training, the model replaced \acrshort{Mutation++} and converged rapidly to a new solution. The newly computed base flows were recovered accurately when compared to the true solution (obtained with \acrshort{Mutation++}). During this demonstration, we observed up to 70 \% CPU time decrease to compute the thermochemical properties of the mixture.
This computational framework can be readily ported into other application fields to accelerate simulations that rely on high-dimensional look-up tables to model complex flow behavior such as combustion, phase-change or fluid-particle interactions. \\
Finally, future steps in algorithmic development will include on-the-fly adaptivity of the model to tackle unsteady flow problems, and leveraging analytical solutions during the initial training set. Application to ablation and more detailed chemical mechanisms, for example those appearing in reactive flows, are also being pursued at the moment in the context of future work.

\section*{Acknowledgments}

This work was supported by the Imperial College London - CNRS PhD Joint Program and was granted access to the HPC/AI resources of TGCC under the allocation 2021-A0102B12426 and 2022-A0122B13432 made by GENCI. Part of the calculations were also performed using MeSU computing platform at Sorbonne University. 

%\begin{appendices}

%\section{First appendix}\label{secA1}

%\end{appendices}


%\bibliography{references.bib}
\end{document}

%% file: preamble.tex
\usepackage[utf8]{inputenc}
\usepackage{graphicx}
\usepackage[caption=false]{subfig}
\usepackage{amsmath}
\usepackage[version=4]{mhchem}
\usepackage{siunitx}
\usepackage{longtable,tabularx}
\usepackage{upgreek}
\usepackage{siunitx}
\usepackage{mathtools}
\usepackage{xcolor}
\usepackage{tikz}
\usepackage{hyperref}

\usepackage[capitalise,nameinlink,sort]{cleveref}
\crefrangeformat{equation}{#3Eqs.~(#1--#2)#6}
\crefname{paragraph}{Section}{Sections}
\Crefname{paragraph}{Section}{Sections}

\input{symbols}

\input{nomenclature}

\glsdisablehyper
\hypersetup{colorlinks=true,linkcolor=blue,citecolor=blue,urlcolor=black}

%% file: symbols.tex
\usepackage{bm}

\newcommand{\pd}[2]{\ensuremath{\frac{\partial #1}{\partial #2}}}

\renewcommand{\vec}[1]{\ensuremath{\mathbf{#1}}}
\newcommand{\newvec}[1]{\ensuremath{\mathbf{#1}}}
\newcommand{\tens}[1]{\ensuremath{\mathrm{#1}}}

\newcommand{\subs}[1]{\ensuremath{\mathrm{#1}}}

\newcommand{\grad}[1]{\ensuremath{\vec{\nabla}#1}}

\newcommand{\divg}[1]{\ensuremath{\vec{\nabla}\cdot#1}}
\newcommand{\divgp}[1]{\ensuremath{\vec{\nabla}\cdot \left( #1 \right)}}

\newcommand{\abs}[1]{\ensuremath{\lvert #1 \lvert}}

\newcommand{\avg}[1]{\ensuremath{\bar{#1}}}
\newcommand{\lin}[1]{\ensuremath{#1}}

\newcommand{\transposep}[1]{\ensuremath{\left({#1}\right)^\mathrm{T}}}

\newcommand{\dimval}[1]{\ensuremath{\tilde{#1}}}
\newcommand{\freestream}[1]{\ensuremath{\dimval{#1}_\freest}}
\newcommand{\valref}[1]{\ensuremath{\dimval{#1}_\subsref}}

\newcommand{\mathset}[1]{\ensuremath{\mathcal{#1}}}

\newcommand{\sumspecies}{\sum_{\speciess\in\speciesset}}
\newcommand{\sumreactions}{\sum_{\reactionr\in\reactionsset}}
\newcommand{\eyematrix}{\tens{I}}

%% file: nomenclature.tex
\usepackage[section=section,nogroupskip,nonumberlist,acronym,nopostdot,hyperfirst=true]{glossaries}
\newglossarystyle{longRaggedRight2cm}{%
 \setglossarystyle{long}%
 \renewenvironment{theglossary}%
    {\begin{longtable}{p{2.5cm}p{\textwidth minus 3cm}}}%
    {\end{longtable}}%
}
\newglossarystyle{longRaggedRight1cm}{%
 \setglossarystyle{long}%
    {\begin{longtable}{p{1cm}p{\textwidth minus 1.5cm}}}%
    {\end{longtable}}%
}

\newglossary[clg]{const}{cns}{con}{Constants}
\newglossary[rlg]{roman}{rom}{rtn}{Roman symbols}
\newglossary[slg]{subsc}{sub}{stn}{Subscripts}
\newglossary[glg]{greek}{grk}{grn}{Greek symbols}
\makeglossaries

\newcommand{\linkvartoglossary}[4]{
	\newglossaryentry{#2}{type=#1,name={\ensuremath{#3}},description={#4},sort={#3}}%
    \expandafter\newcommand\csname #2\endcsname{{\gls{#2}}}%
    \expandafter\newcommand\csname #2i\endcsname[1]{{\gls{#2}[_##1]}}%
    \expandafter\newcommand\csname #2s\endcsname[1]{{\gls{#2}[^##1]}}%
}

\newcommand{\addconst}[3]{\linkvartoglossary{const}{#1}{#2}{#3}}
\newcommand{\addroman}[3]{\linkvartoglossary{roman}{#1}{#2}{#3}}
\newcommand{\addsubsc}[3]{\linkvartoglossary{subsc}{#1}{#2}{#3}}
\newcommand{\addgreek}[3]{\linkvartoglossary{greek}{#1}{#2}{#3}}
\newcommand{\addacronym}[2]{
	\newacronym{#1}{#1}{#2}%
	\expandafter\newcommand\csname #1\endcsname{{\gls{#1}}}%
}
\newcommand{\adddualacronym}[3]{
	\newacronym[description=#3]{#1}{#1}{#2}%
	\expandafter\newcommand\csname #1\endcsname{{\gls{#1}}}%
}

\addconst{RU}{\mathcal{R}_\subs{u}}{}

\addsubsc{speciess}{\subs{s}}{quantity refers to species $\mathrm{s}$}
\addsubsc{directioni}{\subs{i}}{quantity refers to direction $\mathrm{i}$}
\addsubsc{totalnot}{\subs{0}}{quantity refers to total or stagnation (static plus dynamic)}
\addsubsc{freest}{\subs{\infty}}{quantity refers to the free stream at the inlet}
\addsubsc{subsref}{\subs{ref}}{quantity refers to the free stream at the inlet}

\addroman{soundspeed}{c}{speed of sound}
\addroman{pressure}{p}{pressure}
\addroman{temperature}{T}{temperature}
\addroman{mw}{M}{}
\addroman{mwspecies}{\mw_\speciess}{}

\addroman{numspecies}{N_\speciess}{number of species}

\addroman{Eckert}{Ec}{Eckert number at free stream, at inlet, \Eckert = }
\addroman{Mach}{Ma}{Mach number at free stream, at inlet, \Mach = }
\addroman{Reynolds}{Re}{Reynolds number at free stream, at inlet, 
$\Reynolds = \freestream{\density} \freestream{\soundspeed} \valuereference{L} / \valuereference{\viscosity}$}
\addroman{Prandtl}{Pr}{Prandtl number at free stream, at inlet, \Prandtl = }

\addroman{viscosity}{\mu}{viscosity}
\addroman{thermconductivity}{\kappa}{thermal conductivity}
\addroman{heatp}{c_p}{thermal conductivity}

\addroman{internale}{e}{internal specific energy}
\addroman{internalh}{h}{internal specific enthalpy}
\addroman{totale}{\internale_\totalnot}{total specific energy (internal and kinetic)}
\addroman{totalh}{\internalh_\totalnot}{total specific enthalpy (internal and kinetic)}

\addroman{speciese}{\internale_\speciess}{internal specific energy of species $\speciess$}
\addroman{speciesh}{\internalh_\speciess}{internal specific enthalpy of species $\speciess$}

\addroman{speciesmassfrac}{Y_\speciess}{mass fraction of species $\speciess$}
\addroman{speciesmolefrac}{X_\speciess}{mole fraction of species $\speciess$}

\addroman{timevar}{t}{time variable}
\addroman{velocity}{\vec{u}}{hydrodynamic velocity (vector)}
\addroman{veli}{u_i}{hydrodynamic velocity in direction $\directioni$ (scalar)}
\addroman{diffvelocity}{\vec{V}_\speciess}{diffusion velocity for species $\speciess$}

\addroman{heatflux}{\vec{q}}{heat flux vector}

\addroman{statevec}{\newvec{Q}}{}
\addroman{statevecavg}{\avg{\newvec{Q}}}{}
\addroman{stateveclin}{\lin{\newvec{q}}}{}
\addroman{stateveclinhat}{\hat{\stateveclin}}{}

\addroman{fluxvec}{\vec{F}}{}
\addroman{fluxvecconv}{\vec{F}_\subs{C}}{}
\addroman{fluxvecdiff}{\vec{F}_\subs{D}}{}

\addroman{sourcevec}{\vec{S}}{}

\addroman{jacobianA}{\tens{A}}{}
\addroman{jacobianAhat}{\hat{\jacobianA}}{}
\addroman{hessH}{\tens{H}}{}
\addroman{hessHhat}{\hat{\tens{H}}}{}
\addroman{jacobian}{\pd{\fluxvec}{\statevec}}{}

\addroman{resolventR}{\tens{R}_\subs{\jacobianA}}{}
\addroman{svddiag}{\tens{\Sigma}}{}
\addroman{svdin}{\tens{V}}{}
\addroman{svdout}{\tens{U}}{}

\addroman{speciesset}{\mathset{S}}{}
\addroman{reactionsset}{\mathset{R}}{}

\addroman{reactionr}{\subs{r}}{}
\addroman{reactionrate}{R_\reactionr}{}
\addroman{forwardrate}{{k_f}_\reactionr}{}
\addroman{backwardrate}{{k_b}_\reactionr}{}
\addroman{equilconst}{{K_\subs{eq}}_\reactionr}{}

\addroman{forcing}{\vec{f}}{}

\addroman{spacevec}{\vec{x}}{}

\addgreek{massprod}{{\dot{\omega}}_\speciess}{mass production rate due to chemical reactions}
\addgreek{density}{\rho}{density}
\addgreek{densityspecies}{\density_\speciess}{density of species $\speciess$}
\addgreek{viscstresstensor}{\tens{\uptau}}{viscous stress tensor}

\addgreek{gammaratio}{\gamma}{specific heat ratio}

\addgreek{sponge}{\sigma}{}

\adddualacronym{2D}{two-dimensional}{Two-dimensional}
\adddualacronym{3D}{three-dimensional}{Three-dimensional}

\addacronym{AIAA}{American Institute of Aeronautics and Astronautics}

\addacronym{AFOSR}{Air Force Office for Scientific Research}
\addacronym{EOARD}{European Office of Aerospace Research and Development}

\adddualacronym{CFD}{computational fluid dynamics}{Computational Fluid Dynamics}
\addacronym{CFL}{Courant-Friedrichs-Lewy}
\adddualacronym{CNEQ}{chemical non-equilibrium}{Chemical Non-Equilibrium}
\adddualacronym{CPG}{calorically perfect gas}{Calorically Perfect Gas (often called ideal or perfect gas)}

\adddualacronym{DMD}{dynamic mode decomposition}{Dynamic Mode Decomposition}
\adddualacronym{DNS}{direct numerical simulation}{Direct Numerical Simulation}
\adddualacronym{DPLR}{data-parallel line relaxation}{Data-Parallel Line Relaxation (numerical method or computational code)}

\adddualacronym{IBM}{immersed boundary method}{Immersed Boundary Method}

\adddualacronym{LES}{large-eddy simulation}{Large-Eddy Simulation}
\adddualacronym{LST}{linear stability theory}{Linear Stability Theory}
\adddualacronym{LTE}{local thermodynamic equilibrium}{Local Thermodynamic Equilibrium}

\adddualacronym{MFP}{mean-flow perturbation}{Mean-Flow Perturbation}

\adddualacronym{MT}{multi-temperature}{Multi-Temperature}
\adddualacronym{MUTATION}{MUTATION}{MUlticomponent Thermodynamic And Transport properties for IONized gases (library)}
\adddualacronym{Mutation++}{Mutation++}{Multicomponent thermodynamic and transport properties for ionized gases in C++ (library)}

\addacronym{NASA}{National Aeronautics and Space Administration}
\adddualacronym{NEQ}{non-equilibrium}{Non-Equilibrium}

\adddualacronym{PSE}{parabolized stability equations}{Parabolized Stability Equations}

\adddualacronym{SVD}{singular value decomposition}{Singular Value Decomposition}

\adddualacronym{TCNEQ}{thermal and chemical non-equilbirium}{Thermal and Chemical Non-Equilibrium}
\adddualacronym{TNEQ}{thermal non-equilibrium}{Thermal Non-Equilibrium}
\adddualacronym{TPG}{thermally perfect gas}{Thermally Perfect Gas (often called imperfect gas)}
\adddualacronym{TPS}{thermal protection system}{Thermal Protection System}

\adddualacronym{US3D}{{US3D}}{{US3D} code{,} developed by University of Minnesota and NASA}

\adddualacronym{VKI}{von Karman Institute for Fluid Dynamics}{von Karman Institute for Fluid Dynamics (Belgium)}